\documentclass[pre,aps,superscriptaddress]{revtex4} 

\usepackage{color,psfig,graphics,psfrag,amsmath,amssymb,amsfonts,rotating}

\begin{document}

\title{Brillouin and Boson peaks in Glasses from vector Euclidean
Random Matrix Theory} 
\author{S.~Ciliberti} 
\affiliation{INFM UdR
Roma1, Universit\'a di Roma ``La Sapienza'', and Center for
Statistical Mechanics and Complexity (SMC), P.le A. Moro 2, I-00185
Roma, Italy} 
\author{T.~S.~Grigera} 
\affiliation{Centro di Studi e
Ricerche ``Enrico Fermi'', via Panisperna 89/A, I-00184 Roma, Italy}
\author{V.~Mart\'{\i}n-Mayor} 
\affiliation {Departamento de Fisica
Teorica I, Universidad Complutense de Madrid, Madrid 28040, Spain}
\affiliation{Instituto de Biocomputaci\'on y F\'{\i}sica de
Sistemas Complejos (BIFI). Universidad de Zaragoza, 50009 Zaragoza,
Spain.}
\author{G.~Parisi} 
\affiliation{INFM UdR Roma1, Universit\'a di Roma
``La Sapienza'', and Center for Statistical Mechanics and Complexity
(SMC), P.le A. Moro 2, I-00185 Roma, Italy} 
\author{P.~Verrocchio}
\affiliation {Departamento de Fisica Teorica I, Universidad
Complutense de Madrid, Madrid 28040, Spain}
\affiliation{Instituto de Biocomputaci\'on y F\'{\i}sica de
Sistemas Complejos (BIFI). Universidad de Zaragoza, 50009 Zaragoza,
Spain.}

\begin{abstract}
A simple model of harmonic vibrations in topologically disordered
systems, such as glasses and supercooled liquids, is studied
analytically by extending Euclidean Random Matrix Theory to include
vector vibrations. Rather generally, it is found that i) the dynamic
structure factor shows sound-like Brillouin peaks whose
longitudinal/transverse character can only be distinguished for small
transferred momentum, $p$, ii) the model presents a mechanical
instability transition at small densities, for which scaling laws are
analytically predicted and confirmed numerically, iii) the Brillouin
peaks persist deep into the unstable phase, the phase transition being
noticeable mostly in their linewidth, iv) the Brillouin linewidth
scales like $p^2$ in the stable phase, and like $p$ in the unstable
one. The analytical results are checked numerically for a simple
potential. The main features of glassy vibrations previously deduced
from scalar ERMT are not substantially altered by these new results.
\end{abstract}

\maketitle

\section{Introduction}

Recent advances in $X$-ray and neutron scattering techniques have
allowed to obtain very detailed physical insight into the
high-frequency (0.1--10$\,$THz) vibrational dynamics of supercooled
liquids and glasses. Indeed, within this range of frequencies their
spectra reveal several universal properties~\cite{exp}, related with
the presence of sound-like excitations for momenta $p$ of the same
order of magnitude of $p_0$, the first maximum of the static structure
factor (typically corresponding to wavenumbers of a few $nm^{-1}$).
In $X$-ray inelastic scattering experiments, this \emph{high-frequency
sound} is mainly revealed as Brillouin-like peaks in the Thz region of
the dynamic structure factor, whose position grows linearly for $p
<p_0$, i.e. $\omega(p) \sim c p$ (the speed of sound $c$ being quite
close to that obtained by acoustic measurements) and saturates at a
frequency $\omega(p) \sim \omega_0$ for $p \sim p_0$. Moreover, the
$p$-dependence of the peak width is often described by
$\Gamma(p)={\cal A} p^\alpha$, where ${\cal A}$ is basically
temperature independent for momenta ranging from $0.01 p_0$ to
$p_0$~\cite{scopigno03}. Interestingly enough, $\Gamma(p)$ also
saturates as the momentum becomes $\sim p_0$. The actual value of the
exponent $\alpha$ is still a matter of debate among different
experimental groups~\cite{debate} (some proposing $\alpha=2$, some
$\alpha=4$), yet one should bear in mind that for somehow smaller
momenta, an unambiguous $p^2$ scaling has been found in optical
measurements of sound attenuation in amorphous silica~\cite{ZHU91}.

The fact that the above described features, as well as the Boson peak
(see below), are universal, supports the hope that most of the
underlying relevant physics can be captured by some simple model. Yet,
the very nature of these systems, intermediate between solids and
liquids, poses a considerable challenge to the description of their
spectra. One could, for instance, take the liquid point of view: the
well-known \emph{hydrodynamic approximation}~\cite{hansen} predicts
the existence of sound-like excitations for all wave numbers, and a
sound attenuation coefficient (very much the same as the line-width of
a Brillouin peak~\cite{exp}) that grows like $p^2$. And in fact, as
said above, such a scaling has been measured for instance in the range
$0.01 p_0$--$0.1 p_0$ in amorphous silica~\cite{ZHU91}. However, when
using the value of the viscosity of silica at room temperature (an
unknown quantity not smaller than $10^{13}$ poises) in the
hydrodynamic formulas~\cite{hansen,Ribeiro98}, is overestimated by (at
least) eight orders of magnitude. Thus, hydrodynamics predicts a
completely washed-out Brillouin peak, in plain contradiction with
experiments.  A drastically different approach is to consider these
excitations (whose inverse frequency is much smaller than the
structural relaxation time) as harmonic vibrations around a quenched
atomic structure, a point of view supported by recent molecular
dynamics simulations~\cite{ruocco00,horbach99,pilla02}. Given the
presence of well formed local structures (SiO$_2$ tetrahedra, for
instance) a most natural approximation in this context is to consider
that the oscillation centers form a crystalline structure, the
disorder in the atomic positions being mimicked by randomness in their
interaction potential~\cite{schirmacher98,taraskin} (disordered
lattice models~\cite{elliott}). This approximation is particularly
appealing for analyzing scattering experiments, since
inelastic-scattering from crystals is nowadays a well developed
discipline. The presence of more than one atomic species in most glass
formers produces a complicated vibrational structure, with acoustical
and optical branches (both longitudinal and transversal) which are
degenerate in energy for wave numbers close to $p_0$. It is clear that
a crystalline analogue can be very useful to clean-up the
mess~\cite{taraskin}. However, disordered lattice models dramatically
underestimate the scattering of sound waves~\cite{mayor00}.

A somehow intermediate position is held by those studying vibrations
around a topologically disordered~\cite{elliott} (liquid like)
structure. There are basically two such approaches: modified
mode-coupling theory~\cite{mct} (which is not limited to harmonic
excitations), and euclidean random matrix theory
(ERMT)~\cite{ermZee,erm2,erm2bis}. ERMT owes its name to the fact that
it formulates the vibrational problem as random matrix
problem~\cite{metha}. The matrices involved are called Euclidean
random matrices~\cite{erm}, and their study has required the
development of new analytical tools.  Both MCT and ERMT predict an
enhanced scattering of sound waves as compared to disordered crystals,
but up to now had been limited (due to technical difficulties) to very
simple approximations where the three-dimensional nature of particle
vibrations was neglected. This paper deals with the extension of ERMT
to the physical case of three-dimensional vibrations. In this way, we
shall learn that indeed the universal features of the high-frequency
spectra can be ascribed to the topologically disordered structure of
supercooled liquids and glasses. However, we must stress that no
results are available yet for the case were several atomic species are
present with very strong {\em local} correlations among them ({\em
e.g.\/} silica tetrahedrons), which would be directly relevant for the
interpretation of scattering data.

In the previous scalar studies, the line width was found to be
$\Gamma(p) \sim {\cal C} p^2 + {\cal D} p^2 \omega^2$, and the peak
position $\omega=c p$. So for $p \ll p_0$ one gets $\alpha=2$, thus
recovering the hydrodynamic scaling, but with Brillouin peaks that are
still observable when $p$ is close to $p_0$~\cite{erm2bis}. It must be
noticed that the possibility remains of a crossover at a momentum $p_c
< p_0$ (before the broadening saturates) to the $\alpha=4$ regime.
This possibility should be always considered when discussing
experimental results obtained under different thermodynamic conditions
and for different momentum ranges~\cite{debate}.

Another quantity accessible to experiments (especially Raman
scattering) is the vibrational density of states (VDOS), $g(\omega)$.
Its most striking feature is the presence of an excess of states over
the Debye $\omega^2$ law in the ``low'' frequency region, i.e.\ where
the dispersion relation is linear (but still in the Thz region). This
excess of states is seen as a peak when the VDOS is suitably plotted,
and has been named {\em Boson peak} (BP).  There are at least three
different ways of defining the boson peak from experiments. It is
sometimes defined as a peak in Raman scattering data, sometimes as a
peak in the difference between the observed VDOS of the glass and that
of the corresponding crystal (which is $\propto \omega^2$). The third
definition, and the one we will adopt, is to look for a peak in
$g(\omega)/\omega^2$. When so defined, the peak position $\omega_{BP}$
usually shifts to lower frequency on heating~\cite{bpvsheat}, except
for the case of silica~\cite{bpsilica1}. In this material the shift is
seen on lowering the density~\cite{bpsilica2}.

Though there are theoretical approaches that explain the BP through
anharmonic effects~\cite{soft-potential}, there is a growing consensus
that anharmonicity, although certainly present in real materials, need
not be invoked to explain the BP or the other features of high
frequency sound described above. Indeed, several numerical simulations
have shown that a model of harmonic vibrations is wholly adequate to
describe this frequency range~\cite{ruocco00,horbach99,pilla02}. On
the other hand, even within the harmonic framework the nature of the
extra low frequency modes giving rise to the BP is still an open
point. At a qualitative level, the frequency $\omega_{BP}$ is close to
the Ioffe-Regel frequency $\omega_{IR}$, the frequency at which
vibrational modes change from propagating to non-propagating. This is
accompanied by a crossover from weak to strong scattering of the
phonons by the disorder~\cite{taraskin,iofferegel}, suggesting the
possibility that the excess BP modes are
localized~\cite{alexander89,courtens96}. However, numerical
simulations of amorphous silica have shown that the localization edge
is at frequencies greater than $\omega_{BP}$ and
$\omega_{IR}$~\cite{diffusoni}. This has also been found analytically
in the CPA framework~\cite{taraskin}. What the Ioffe-Regel criterion
signals is rather a crossover from phonon-like excitations to a
different region where the scattering due to disorder is very strong
and the modes do not propagate. We call these modes {\em glassons}
(since they do not propagate but ``diffuse'', they have also been
called \emph{diffusons}~\cite{diffusoni}). A large bump of glassons is
generally found around the Ioffe-Regel frequency, due to the
flattening of the dispersion relation. This can be considered as the
glass counterpart of the van Hove singularity of
crystals~\cite{taraskin,erm3,pilla02}. All the recently proposed
theoretical frameworks predict that this peak of glassons should move
to lower frequencies when approaching an instability transition, where
negative eigenvalues (imaginary frequencies) appear. This has been
related to the arising of the
BP~\cite{schirmacher98,mct,taraskin,erm3}. Some recent simulations for
silica have stressed that the BP has an strong component of transverse
modes~\cite{pilla02,scopigno02}.

A recent breakthrough of the ERM theory has been to push the analysis
of the vibrational spectra of glasses from the qualitative level,
where mainly orders of magnitude are compared, to a more quantitative
one, making sharp predictions about the values of universal critical
exponents describing the approach to the
singularity~\cite{nature}. The first attempt to confirm the theory by
measuring them in a numeric simulation has been quite
encouraging~\cite{nature}. The emerging picture is the following: the
BP modes are given by the hybridization between the phonons and the
low-energy tail of the glasson peak which softens when the system
approaches the instability transition.  Such a mechanism is strongly
suspected to take place in supercooled liquids at the Mode Coupling
transition, and there is numerical support \cite{saddles} for the idea
that the Mode Coupling transition (actually, cross-over) in
supercooled liquids corresponds to a smooth phase transition in the
energy landscape, from a saddle-dominated region to a minima-dominated
one.

Interestingly enough, since the ERM computation was originally
performed in the monoatomic scalar case (i.e. the vibrations are all
collinear) the results obtained do not depend on the existence of
transversal or optic modes but describe a general phenomenon occurring
when phonons and glassons interact. On the other hand, since in real
systems the lower energy glassons have been
claimed~\cite{pilla02,scopigno02} to have transverse polarization, in
this work we shall support the universal character of the transition
by extending the Euclidean Random Matrix Theory to a generic model
with longitudinal and transverse modes. Our aim is to check the
validity of the predictions of the vectorial ERM computation on a
simple Gaussian model whose spectral properties have been numerically
studied by the method of moments~\cite{mom}. As a matter of fact the
theory predicts that the behaviour of some of the main spectral
features, namely the arising of the BP and the broadening of the
Brillouin peak, are universal and hence can be captured even by the
simplest model. We shall show concerning the features mentioned above
that the numeric results of the Gaussian model agree with the
theoretical predictions and are identical to those found in more
realistic models of fragile glass formers~\cite{nature,ruocco00}.

The layout of the rest of this paper is as follows. In section~II we
introduce the theoretical formalism of ERM theory. In the next section
we discuss the phase transition as it results from an analytical
model-independent computation. In section IV we check these
conclusions by reporting numerical and analytical results of a
particular simplified model (the Gaussian model). We will see that our
analytical results fail in giving the whole shape of the spectrum, due
to the superposition approximation and to the high density
expansion. However, these results describe rather well the behaviour
of the system near the transition point, with the correct exponents
and scaling laws. We summarize our conclusions in the last section.

\section{The Euclidean Random Matrix Theory in the vector case}

We study a model where particles oscillate around fixed random
positions, so that the position of particle $i$ at time $t$ is ${\bf
x}_i(t) = {\bf x}^{eq}_i + {\bf \varphi}_i(t)$; the ${\bf x}^{eq}_i$
are quenched equilibrium positions (whose distribution must be
specified) and ${\bf \varphi}_i(t)$ are the displacements.  From now
on, greek indices will label the Cartesian components of the
displacements, ${\bf \varphi}_i(t)$.  In the harmonic approximation
the Hamiltonian is
\begin{equation} 
H\left[{\bf x}\right]= \sum_{i,j}^{1,N}v({\bf x}_i-{\bf x}_j) \simeq
\frac 12\sum_{i,j}^{1,N}\sum_{\mu,\nu}^{1,3} M_{i\mu,j\nu}[{\bf
x}^{eq}]\varphi_i^\mu\varphi_j^\nu
\end{equation} 
where the dynamical matrix M is an Euclidean Random Matrix:
\begin{equation} 
M_{i\mu,j\nu}[{\bf x}^{eq}]\equiv
- f_{\mu\nu}({\bf x}_i^{eq}-{\bf x}_j^{eq})
\ +\ 
\delta_{ij}\sum_{k=1}^N f_{\mu\nu}({\bf x}_i^{eq}- {\bf x}_k^{eq}),
\end{equation} 
with $f_{\mu\nu}({\bf x})\equiv\partial_{\mu\nu} v({\bf
x})$. Translational invariance implies that there are 3 null
eigenvalues corresponding to the rigid translations of the system as a
whole.

In the one-excitation approximation (and in the classical limit) the
dynamic structure factor measured in inelastic-scattering experiments
is
\begin{equation} 
S^{(1)}({\bf p},\omega)= \frac{k_{\mathrm B} T}{m\omega^2} \overline{
\sum_n \left| \sum_i {\bf p}\cdot{\bf e}_{n,i} e^{ i{\bf p} \cdot {\bf
x}^{eq}_i } \right|^2 \delta(\omega -\omega_n)},
\end{equation} 
where ${\bf e}_n$ are the eigenvectors of the dynamical matrix and
$\omega_n$ its eigenfrequencies (= square root of eigenvalues).
The overline means average over the disordered quenched positions,
whose distribution $P[ {\bf x}^{eq} ]$ has to be specified. It is thus
assumed (as it is often the case for disordered systems) that
macroscopic observables are self-averaging.  The density of states
(VDOS) is obtained in the limit of large momenta:
\begin{equation}
g(\omega)=\lim_{p\to\infty} {m\omega^2 \over k_B T p^2} 
S^{(1)}(p,\omega).   
\label{s-dos}
\end{equation}
We can obtain $S^{(1)}({\bf p},\omega)$ through the resolvent 
$G({\bf p},z)$:
\begin{equation}
G_{\mu\nu}({\bf p},z) \equiv \frac{1}{N} \sum_{jk}
\overline{ 
e^{i {\bf p} \cdot (
{\bf x }^{\mathrm eq}_j-{\bf x}^{eq}_k ) }
\left[ {1 \over z-M} \right]}_{j\mu,k\nu} 
\equiv G_L(p,z)\frac{ p_\mu p_\nu}{p^2} + 
G_T(p,z)\left(\delta_{\mu\nu}-\frac{ p_\mu p_\nu}{p^2}\right)  \label{s-res}
\end{equation}
which is an axial tensor that can be separated in a longitudinal term
and a transversal one, $G_L(p)$ and $G_T(p)$, depending only on the
magnitude of {\bf p}.  The dynamic structure factor is obtained from
the longitudinal resolvent using the distribution identity
$(x+i0^+)^{-1} = P(1/x) -i\pi\delta(x)$ (Plemelj formula):
\begin{equation}
S^{(1)}({\bf p},\omega)= 
- \frac{2 k_\mathrm{B} T p^2}{\omega\pi} 
\mathrm{Im}\, G_{L}({\bf p},\omega^2 +{\mathrm i} 0^+).
\label{sdiq}
\end{equation}
A transverse dynamic structure factor (not measurable in experiments)
can be defined in an analogous way, and will have a Brillouin peak 
corresponding to the transverse excitations. 

A most important and general result is that for $p\to\infty$ the
resolvent becomes isotropic:
\begin{equation}
G^\infty_{\mu\nu}(z) = \frac{1}{N} \sum_{j}
\overline{
\left[ {1 \over z-M} \right]}_{j\mu,j\nu} 
= \delta_{\mu\nu} \frac 1N \overline{ \textrm{Tr}\: [z-M]^{-1} }.
\end{equation}
So both longitudinal and transverse structure factors tend to a common
limit (the VDOS, see eq.~\ref{s-dos}) at infinite momentum.  This
implies in principle that both the dispersion relations saturate at
the same value. However in disordered systems the dispersion relation
is ill-defined when $\omega \sim \omega_{IR}$ since the Brillouin peak
becomes very broad. We shall come back to this subject later.

In order to fully define the model one has to specify the potential
$v(r)$, which for the moment we leave unspecified, and the probability
distribution $P[{\bf x}^{eq}]$. The choice of $P$ is a crucial one,
the various alternatives having a different physical meaning. Two
cases are especially interesting in the physics of glasses. If one is
interested in the instantaneous normal mode (INM) spectrum, one should
choose
\begin{equation}
P[{\bf x}^{eq}]\propto \exp{\left(-\beta H[{\bf x}^{eq}]\right)} ,
\label{p-inm}
\end{equation}
with $\beta=1/{k_B T}$, since one is studying the vibrations around
typical equilibrium configuration~\cite{INM}. If instead one wants to
focus on vibrations around stationary points (saddles or minima) of
the Hamiltonian with a given energy $E$, the choice would be
\begin{equation}
P[{\bf x}^{eq},E] \equiv \frac{1}{{\cal N}(E)} \sum_{\alpha}
\delta\left[ {\bf x}_1-{\bf x}^{\alpha}_1(E) \right] \ldots
\delta\left[ {\bf x}_N-{\bf x}^{\alpha}_N(E) \right] ,
\label{p-stat}
\end{equation}
where $\{{\bf x}^{\alpha}_i(E)\}$ are the ${\cal N}(E)$ solutions of
the equations $\partial_i H({\bf x})=0$ under the condition $H({\bf
x}^{\alpha}_i)=E$.  The probability function enters in the calculation
of integrals of the type
\begin{equation}
\int d{\bf x}_1 d{\bf x}_2\ldots d{\bf x}_k g^{(k)}({\bf x}_1,{\bf
x}_2,\ldots{\bf x}_k) f({\bf x}_1,{\bf x}_2)f({\bf x}_2,{\bf
x}_3)\ldots f({\bf x}_{k-1},{\bf x}_k),
\label{integrale}
\end{equation}
where $g^{(k)}({\bf x}_1,{\bf x}_2,\ldots{\bf x}_k) = \rho^{-k} \int
\!\! d{\bf x}_{k+1}\ldots d{\bf x}_N\, P[{\bf x}]$ is the $k$-particle
distribution function. Obviously the computations become a lot easier
by taking $P[{\bf x}]=1/V^N$ (i.e.\ considering an uniform
distribution of the equilibriom positions), instead of the actual
forms (\ref{p-inm}) or (\ref{p-stat}). This is what we shall do in the
following. As we shall later discuss, this brutal approximation is not
completely useless, as might seem at first sight. For the moment, let
us show the main steps of the computation in this simplified
situation.

The computation of (\ref{s-res}) is the generalization of the scalar
vibrations case \cite{erm2,erm2bis}, so we shall only outline it,
stressing the differences.  Proceeding as in the scalar case, one
first expands the propagator in $1/z$:
\begin{equation}
G_{\mu\nu}({\bf p},z)= \sum_{r=0}^\infty \frac{1}{z^{r+1}}
\frac{1}{N} \sum_{jk}
\overline{ 
e^{i {\bf p} \cdot (
{\bf x }^{eq}_j-{\bf x}^{eq}_k ) }  (M^r)_{j\mu,k\nu}}.
\label{OOZ}
\end{equation}
Consider the term of order $r$ in this expansion. If we do not allow
repetitions, there are $N!/(N-r-1)!\approx N^{r+1}$ ways of choosing
the particle index. There is a factor $1/V^{r+1}$ from the average
over particle positions, another factor $1/N$ from the definition of
the resolvent and an extra factor of $V$ due to translational
invariance. Hence terms without repetitions contribute $O(\rho^r)$
($\rho=N/V$ is the particle number density). It is easy to convince
oneself that every time that we allow a particle repetition, we loose
a factor of $\rho$. Then, the generic term in the $1/z$ expansion is a
polynomial in $\rho$,
\begin{equation}
\frac{1}{N} \sum_{jk}
\overline{ 
e^{i {\bf p} \cdot (
{\bf x }^{eq}_j-{\bf x}^{eq}_k ) }  (M^r)_{j\mu,k\nu}}=
\rho^r A^{(r,r)}_{\mu\nu}({\bf p})  + 
\rho^{r-1} A^{(r,r-1)}_{\mu \nu}({\bf p})  +\ldots+
\rho A^{(r,1)}_{\mu \nu}({\bf p}) .
\end{equation}
Let us remark that for  $p=0$ we have
\begin{equation}
G_{\mu\nu}({\bf 0},z) = \frac{\delta_{\mu\nu}}{z}, \label{unoz}
\end{equation}
which means that uniform displacements are eigenvectors with zero
eigenvalue. This is a consequence of translational symmetry, and must
hold at arbitrary density. Therefore, we must have 
$A^{(r,s)}_{\mu \nu}(p=0)=0$ for all $r$ and $s$.

The calculation proceeds by finding the polynomial coefficients
$A^{(r,s)}_{\mu \nu}({\bf p})$ up to a given approximation ($1/\rho$
expansion) to all orders $r$. Then the $1/z$ expansion is
resummed. The leading order (that with the highest order in $\rho$)
defines the {\em bare} propagator. The terms with smaller powers of
$\rho$ will take the form of a self-energy, broadening the spectral
line. Organizing the calculation as explained in
ref.~\onlinecite{erm2}, one easily finds $A^{(r,r)}_{\mu \nu}({\bf
p})$ in terms of $\hat f_{\mu \nu}({\bf p})$, the Fourier transform of
$f_{\mu \nu}({\bf r})$, that can be separated in longitudinal
transverse parts:
\begin{equation}
\hat f_{\mu \nu}({\bf p})=\hat f_L(p) \frac{p_\mu p_\nu}{p^2} + \hat
f_T(p) \left(\delta_{\mu \nu} - \frac{p_\mu p_\nu}{p^2} \right).
\end{equation}
In the following it will be important to remember that $\hat
f_L(0)=\hat f_T(0)\equiv\hat f(0)$. One finds
\begin{equation}
 A^{(r,r)}_{\mu \nu}({\bf p}) = \left(\left[\hat f (0) -\hat f({\bf
 p})\right]^r\, \right)_{\mu \nu} = \left[\hat f(0)-\hat
 f_L(p)\right]^r\, \frac{p_\mu p_\nu}{p^2} + \left[\hat f(0)-\hat
 f_T(p)\right]^r\, \left(\delta_{\mu \nu} - \frac{p_\mu p_\nu}{p^2}
 \right)\,.
\end{equation}
Thus, at the leading order in the high-density expansion, the $1/z$
series is geometric and the longitudinal and transverse part are
independent. The {\em bare} propagator (exact in the $\rho\to\infty$
limit) is then
\begin{eqnarray}
G^{(0)}_{\mu \nu}({\bf p},z)&=& G^{(0)}_L(p,z) \frac{p_\mu p_\nu}{p^2} 
+ G^{(0)}_T(p,z)
\left(\delta_{\mu \nu} - \frac{p_\mu p_\nu}{p^2} \right),\\
G^{(0)}_{L,T}({\bf p},z)&=&\left[z- \rho f({\bf 0})+\rho
f_{L,T}({\bf p}) \right]^{-1}.
\end{eqnarray}
Notice that it verifies Eq.(\ref{unoz}), as it should.
Using~(\ref{sdiq}) the Brillouin peaks in the dynamic structure factor
can be recovered in the $\rho\to\infty$ limit. Both the longitudinal
and the transverse propagators have a simple pole for each ${\bf p}$,
implying that sound-waves are exact eigenvectors of the dynamical
matrix in this limit. The bare propagator describes an elastic medium
of infinite density where the longitudinal and transverse dispersion
relation are given by
\begin{equation}
\omega^0_{L,T}({\bf p})\equiv(\rho  \hat f({\bf 0})- 
\rho  \hat f_{L,T}({\bf p}))^{1/2},
\label{nuda}
\end{equation}
which become linear at small $p$. The density of states, that is
obtained from the propagator in the $p\to\infty$ limit, is rather
unnatural for $\rho\to\infty$~\cite{ermZee}: a single Dirac delta
function at frequency $\sqrt{\rho \hat f(0)} $. This pathology is a
consequence of the lack of a Debye cut-off frequency in our elastic
medium (there are infinitely more wave numbers {\em outside} an sphere
of arbitrary large radius, $p_c$, than inside it, but all wave numbers
larger than $p_c$ will have a frequency arbitrarily close to
$\sqrt{\rho \hat f(0)}$). This problem will be cured by the finite
$\rho$ corrections.

\begin{figure}
\centerline{\psfig{figure=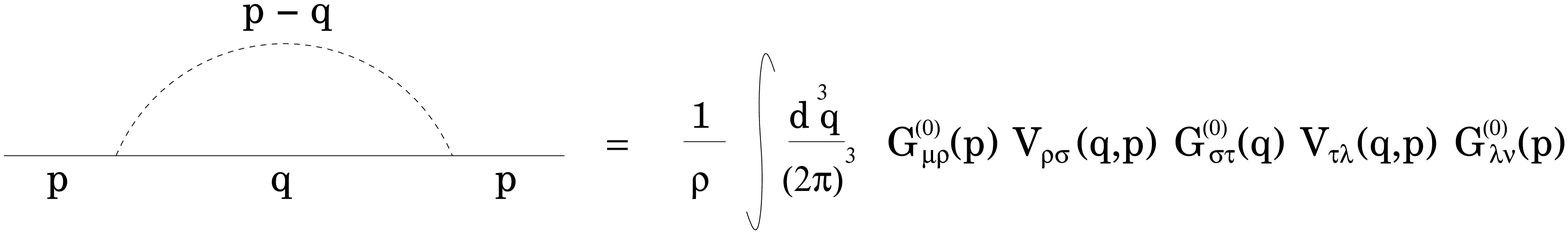,width=\columnwidth,angle=0}}
\caption{\small A simple diagram illustrates the Feynman rules.}
\label{Fig-diagrams}
\end{figure}

\begin{figure}
\centerline{\psfig{figure=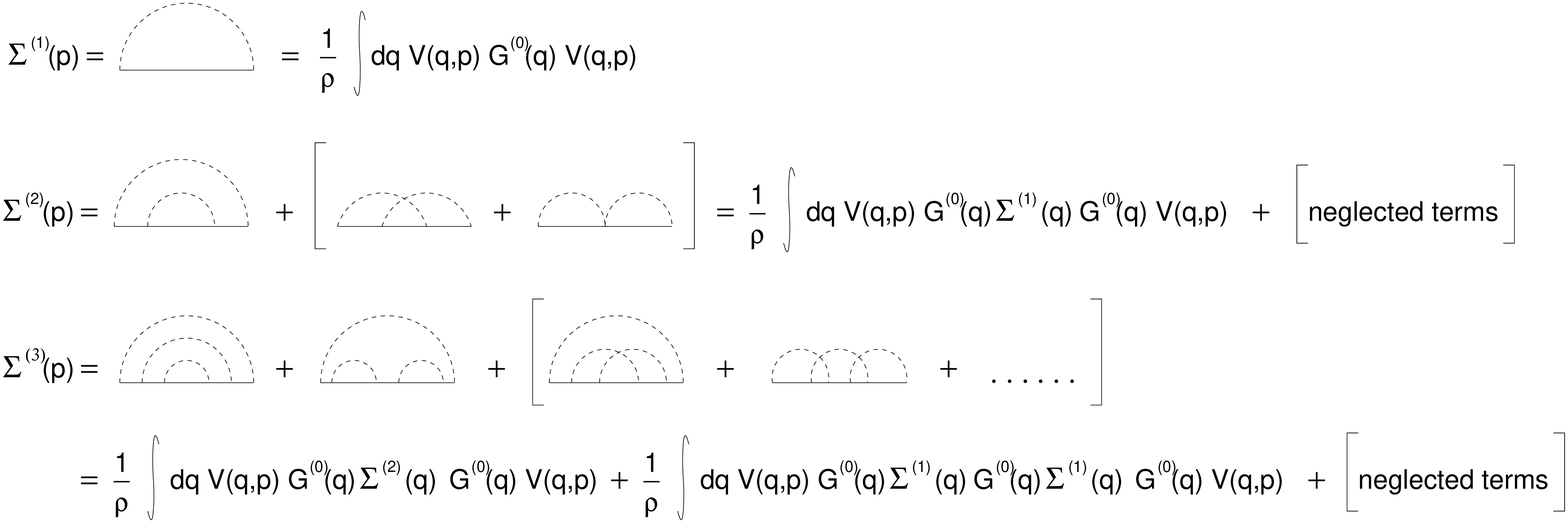,width=\columnwidth,angle=0}}
\caption{\small The diagrams corresponding to the first three terms in
$1/\rho$ of the self-energy $\Sigma({\bf p},z)$ (see
eq.~(\ref{sum-sigma})). Non-planar graphs (shown inside the brackets)
are neglected in the \emph{cactus} resummation of the Dyson equation.}
\label{sigma}
\end{figure}

To calculate the finite $\rho$ corrections, one must allow for
particle label repetitions in the matrix products~(\ref{OOZ}).
Representing the matrix product by a chain of particle
labels, at first order in $1/\rho$ we have to take care of a single
particle label repetition,
\begin{equation}
\ldots  1  \ldots  1 \ldots,
\label{oneloop}
\end{equation}
where the unrepeated indices are represented by dots. Calculating the
generic contribution of this diagrams for all $r$, one obtains the
coefficients $A^{(r,r-1)}_{\mu \nu}({\bf p})$, then, resumming the
$1/z$ expansion, one finds (repeated greek indices are summed):
\begin{eqnarray}
G_{\mu \nu}({\bf p},z)&=& G_{\mu \nu}^{(0)}({\bf p},z)\ +\
G_{\mu \eta}^{(0)}({\bf p},z) \left[\frac{1}{{\rho}} \int \!\! \frac{d^3
q}{(2\pi)^3} V_{\eta\lambda}({\bf q},{\bf p})G^{(0)}_{\lambda\sigma}({\bf
q},z) V_{\sigma\rho}({\bf q},{\bf p})\right]
G_{\rho \nu}^{(0)}({\bf p},z)+ {\cal O}\left(\frac 1{\rho^2}\right),
\label{PIPPO}\\ 
V_{\mu\nu}({\bf q},{\bf p})&=&\rho (\hat
f_{\mu\nu}({\bf q}) - \hat f_{\mu\nu}({\bf p}-{\bf q})).
\end{eqnarray}
This can be easily interpreted if one has in mind that corrections to
the $\rho\to\infty$ limit will take the form of a self-energy:
\begin{equation}
 G_{\mu \nu}({\bf p},z)=\left[\frac{1}{z-\rho \hat f(0)+ \rho \hat
 f({\bf p}) - \Sigma ({\bf p},z)}\right]_{\mu \nu},
\end{equation}
where $\Sigma ({\bf p},z)$ is a matrix with the standard form
\begin{equation}
\Sigma_{\mu\nu}({\bf p},z)=\Sigma_L(p) \frac{p_\mu p_\nu}{p^2} +
\Sigma_T(p) \left(\delta_{\mu \nu} - \frac{p_\mu p_\nu}{p^2}
\right).
\end{equation}
Due to (\ref{unoz}) the self-energy should vanish at $p=0$.
Now, if the self-energy has a series expansion in $1/\rho$,
\begin{equation}
\Sigma ({\bf p},z) = \Sigma^{(1)} ({\bf p},z) + 
\Sigma^{(2)} ({\bf p},z) +\ldots\,
\label{sum-sigma}
\end{equation}
the contribution of order $1/\rho^l$ being $\Sigma^{(l)} ({\bf
p},z)$, then the resolvent reads (Dyson resummation):
\begin{eqnarray}
 G_{\mu \nu}({\bf p},z) & = & G^{(0)}_{\mu \nu}({\bf p},z) +
 G^{(0)}_{\mu \eta}({\bf p},z) \Sigma^{(1)}_{\eta \lambda} ({\bf p},z)
 G^{(0)}_{\lambda \nu}({\bf p},z) + G^{(0)}_{\mu \eta}({\bf p},z)
 \Sigma^{(2)}_{\eta \lambda} ({\bf p},z) G^{(0)}_{\lambda \nu}({\bf
 p},z) +\nonumber\\ 
 && G^{(0)}_{\mu \eta}({\bf p},z)
 \Sigma^{(1)}_{\eta \lambda} ({\bf p},z) G^{(0)}_{\lambda \rho}({\bf
 p},z) \Sigma^{(1)}_{\rho \sigma} ({\bf p},z)
 G^{(0)}_{\sigma \nu}({\bf p},z) + \ldots
\end{eqnarray}
Therefore, one is tempted to conclude from eq.~(\ref{PIPPO}) that
\begin{equation}
\Sigma^{(1)}_{\mu\nu}({\bf p},z)=\frac{1}{{\rho}} \int\frac{d^3 q}{(2\pi)^3} 
V_{\mu\lambda}({\bf q},{\bf p})G^{(0)}_{\lambda\sigma}({\bf q},z)
V_{\sigma\nu}({\bf q},{\bf p}).
\end{equation}
To see that this is actually the case, one can consider the second
order corrections, where we have four different topologies:
\begin{eqnarray}
\ldots 1 \ldots 1 \ldots 2 \ldots 2 \ldots\label{ESQUEMA1},\\ 
\ldots 1 \ldots 2 \ldots 2 \ldots 1 \ldots\label{ESQUEMA2},\\ 
\ldots 1 \ldots 2 \ldots 1 \ldots 2 \ldots\label{ESQUEMA3},\\
\ldots 1 \ldots 1 \ldots 1 \ldots\label{ESQUEMA4}.
\end{eqnarray}
It is clear that to make further progress one needs some kind of
Feynman rules.  We have been able to find a set of diagrammatic rules
valid for planar topologies (i.e.\ repeated indices can be nested but
not intercalated) like eqs.~(\ref{ESQUEMA1}) and~(\ref{ESQUEMA2}),
which we give below. They should not be used in general situations
like eq.(\ref{ESQUEMA4}). The rules are (see Fig.~\ref{Fig-diagrams}):
\begin{enumerate}
\item Join unrepeated particle indices with a full line.
\item Join repeated particle indices with a dashed ``interaction''
line.  Momentum is conserved at each interaction vertex, so we can
attach a momentum to each line.
\item To a full line carrying momentum ${\bf k}$ associate a bare
propagator $G^{(0)}_{\lambda\sigma}({\bf k},z)$.
\item Place a vertex function $V({\bf q},{\bf p})$ at each of the
vertices connected by a dashed line. Here ${\bf p}$ is the incoming
momentum and ${\bf q}$ is the one running inside the loop. Beware the
vertex function does not commute with the bare propagator, so the
order they have in the diagram must be respected. The resulting
sequence must be interpreted as a matrix product.
\item Integrate over the momenta inside the loops. 
\item Add a factor $1/\rho$ for each particle repetition (dashed line).
\end{enumerate}
In this way, one can easily see that eq.~(\ref{ESQUEMA1}) represents a
term of the Dyson resummation of $\Sigma^{(1)}$, while the other three
schemes are genuine contributions to $\Sigma^{(2)}$. These rules
represent a significant improvement over the previously published
expansions \cite{erm2}. For example, the repetition
scheme~(\ref{ESQUEMA2}) involved 16 diagrams in the previous version
(see appendix of ref.~\onlinecite{erm2}), while it now corresponds to
just one.  The other two diagrams for $\Sigma^{(2)}$, which will be
neglected in the resummation used below (see Fig.~\ref{sigma}) are the
diagram corresponding to the particle repetition in~(\ref{ESQUEMA3}),
\begin{eqnarray}
\frac{1}{\rho^2} \int\!\!\frac{d^3q}{(2\pi)^3}\,
\frac{d^3k}{(2\pi)^3}\,
    V_{\mu\alpha}({\bf p}-{\bf q},{\bf p}) G^{(0)}_{\alpha\beta}(z,{\bf q}) 
    V_{\beta\gamma}({\bf q}-{\bf k},{\bf q}) G^{(0)}_{\gamma\delta}(z,{\bf k})
    V_{\delta\sigma}({\bf p}-{\bf q},{\bf p}-{\bf q}+{\bf k}) \times 
    \nonumber\\
     \times G^{(0)}_{\sigma\tau}(z,{\bf p}-{\bf q}+{\bf k}) 
    V_{\tau\nu}({\bf q}-{\bf k},{\bf p}),
\end{eqnarray}
and that corresponding to~(\ref{ESQUEMA4}),
\begin{eqnarray}
\frac{1}{\rho^2} \int\!\!\frac{d^3 q}{(2\pi)^3}\,
\frac{d^3 k}{(2\pi)^3}\, V_{\mu\alpha}({\bf p}-{\bf q},{\bf p})
     G^{(0)}_{\alpha\beta}(z,{\bf q}) V_{\beta\gamma}({\bf q}-{\bf
k},2{\bf q}-{\bf p}) \times \nonumber\\
 \times G^{(0)}_{\gamma\delta}(z,{\bf k}) V_{\delta\nu}({\bf p}-{\bf
k},{\bf p}).
\end{eqnarray}
This three diagrams (see Fig.\ref{sigma}), vanish independently for
${\bf p=0}$, and (in the scalar case) are equivalent to the 39
diagrams of the previuosly published expansion\cite{erm2}.

Given these rules, it is a well-known combinatorial result (Dyson
equation) that the sum of all planar diagrams ({\em cactus}
approximation) takes the form of a self-consistent integral equation
for the self-energy:
\begin{equation} 
\Sigma_{\mu\nu}({\bf p},z)=\frac{1}{{\rho}} \int\frac{d^3 q}{(2\pi)^3} 
V_{\mu\lambda}({\bf q},{\bf p})G_{\lambda\sigma}({\bf q},z)
V_{\sigma\nu}({\bf q},{\bf p}).
\label{eqint}
\end{equation}
The first terms leading to this equation are illustrated in
Fig.~\ref{sigma}. It is most important that this resummation is
compatible with Eq.(\ref{unoz}). An asymptotic model-independent
analysis of this equation will be presented in the next section, while
numerical results for a simple model will be given in section~IV.

Let us remark that the self energy renormalizes the dispersion
relations and gives a finite width to the Brillouin peaks:
\begin{eqnarray}
\omega^2_{L,T}(p)&=&
\left(\omega^0_{L,T}\right)^2 (p)+ \textrm{Re} \:
\Sigma_{L,T}(p,\omega_{L,T}(p)),
\nonumber \\
\Gamma_{L,T}(p)&=&\textrm{Im} \:
\Sigma_{L,T}(p,\omega_{L,T}(p))/\omega_{L,T}(p).
\label{allargamento}
\end{eqnarray}
It is interesting to notice that even if the bare propagator did not
have a transverse component the first order contribution to the
self-energy $\Sigma^{(1)}$ would generate a transverse excitation
band: the scattering among longitudinal and transversal phonons is
much stronger in our case than in disordered lattices.

Before proceeding to the analysis of the cactus equation, let us note
that the correlations between the equilibrium positions of the
particles can be taken into account quite easily at the level of the
{\em superposition approximation} in the above approach. This
approximation amounts to writing the integral~(\ref{integrale}) as
\begin{equation}
\int \!\!
d{\bf x}_1 d{\bf x}_2\ldots d{\bf x}_k \,
g^{(2)}({\bf x}_1,{\bf x}_2)g^{(2)}({\bf x}_2,{\bf x}_3)\ldots
g^{(2)}({\bf x}_{k-1},{\bf x}_k)
f({\bf x}_1,{\bf x}_2)f({\bf x}_2,{\bf x}_3)\ldots
f({\bf x}_{N-1},{\bf x}_k) ,
\label{superposition}
\end{equation}
where all the information about the correlations is assumed to be
contained pair distribution function $g^{(2)}({\bf x}_1,{\bf x}_2)$ 
(corresponding to the chosen probability $P[{\bf x}^{eq}]$). The results
derived above for the case without correlations are translated to the
correlated case (at level of the superposition approximation) by
replacing the functions $f({\bf x})$ by $g^{(2)}({\bf x}) f({\bf
x})$~\cite{workinprogress}. In this way the usual power law divergence
of the pair potential for $|{\bf x}|\to 0$ is balanced by the
exponential behaviour of the pair distribution function, and this
ensures the existence of the Fourier transform of the product $f({\bf
x})g^{(2)}({\bf x})$.

\section{Analytical results: the phase transition}

In this section we aim to show that from the cactus approximation
(equation~\ref{eqint}) it is possible to derive a few analytic
model-independent results about the arising of the Boson Peak and the
broadening of the Brillouin peak. These results are expressed in form
of scaling laws, whose exponents are predicted in this
approximation. As the scaling laws obtained do not not depend on the
details of the interaction, we claim that they are a general feature
of topologically disordered systems. It cannot be excluded, however,
that the values of the exponents depend on the approximation chosen.
Simulations and experiments will allow to clarify this point.

Recalling eq.(\ref{s-dos}), the VDOS can be obtained from
\begin{equation}
g(\omega)=-\frac{2\omega}\pi\ \textrm{Im}\ G^\infty(\omega^2+i0^+) ,
\label{dos}
\end{equation}
setting in our equation $z=\omega^2+i0^+$, where
$G^\infty(z)\equiv\lim_{p\to\infty} G({\bf p},z)$.

In order to obtain the quantity $G^\infty(z)$ one has to solve
the integral equation~(\ref{eqint}) in the $p\to\infty$ limit:
\begin{equation}
\frac{1}{G^\infty(z)} = z - \rho \hat f(0)-
\rho A G^\infty(z)
- \rho\int\!\!\frac{d^3q}{(2\pi)^3}\, \hat f^2({\bf q})G({\bf q},z) 
\label{asint}
\end{equation}
where $G^\infty(z)$, $A \equiv (2\pi)^{-3} \int d^3q \hat f^2({\bf
q})$ and the last term (given by the integral of the product
of two anisotropic matrices) in the above equation are matrices
proportional to the identity since in the infinite momentum limit
everything is isotropic.

In order to deal analytically with (\ref{asint}), the crudest
approximation is to neglect the last term, in which case it becomes
quadratic in $G^{\infty}(z)$, and one easily finds a semicircular
VDOS, with center at $\omega= \rho \hat f(0)$ and radius $2\sqrt{\rho
A}$.  Clearly enough, this approximation misses completely the low
frequency part and it is not suitable to describe the phonons. It
rather describes the non-propagating but extended modes at higher
frequencies (glassons). Though the glassons are qualitatively taken
into account by this approximation, the semicircular shape of this
part of the VDOS is an artifact of the approximation, and as we shall
see later the complete theory yields a shape more similar to that
found in real systems.

By reintroducing in an approximated way the missing term of
(\ref{asint}), we are able to study the region of the small
frequencies. In fact, it is possible to show that it can be written
as
\begin{equation}
\int\!\!\frac{d^3q}{(2\pi)^3}\, \hat f^2({\bf q}) 
G({\bf q},\omega^2+i0^+)\simeq 
-\frac{b}{\rho c^3}-i\frac {\hat f^2(0)\omega}{4\pi c^3} ,
\end{equation}
where $b$ is a positive constant and where the sound velocity $c$ 
is such that 
\begin{equation}
\frac 1 {c^3} = \frac 1 {3c_L^3} + \frac 2{3c_T^3},
\qquad
c_{L,T}=\sqrt{-\rho\hat f_{L,T}^{''}(0)/2}.
\label{c}
\end{equation}
It is not difficult to show that $A=a/c^3$, where $a$ does not contain
$c$. Substituting this in eq.~(\ref{asint}) we obtain a quadratic
equation for $G^\infty(\omega^2+i0^+)$ that gives
\begin{eqnarray}
G^\infty(\omega^2+i0^+)&\simeq&
\left[(\omega^2-\rho\hat f(0) + b/c^3) + 
i \rho\hat f^2(0)\omega/(4\pi c^3)\right.\nonumber\\
&-&
\left.\sqrt{\left((\omega^2-\rho\hat f(0) + b/c^3) + 
i \rho\hat f^2(0)\omega/(4\pi c^3)\right)^2-4\rho a/c^3}\right]
\frac 1 {2\rho a/c^3}
\label{ginf}
\end{eqnarray}
In order to study the Boson peak we shall focus in the small
frequencies regime. For pedagogical purposes, it will very
illuminating to consider the case where the VDOS changes because of
changes in the density. However, the mechanism explaining the arising
of the BP is much more general and can be applied even to case where
the density is kept fixed and other thermodynamic parameters vary.

Let us consider first the high density limit, i.e.\ the situation where
the first term under the square root is much larger than the
second. The resolvent becomes
\begin{equation}
G^\infty(\omega^2+i0^+)\simeq 
\left(\omega^2-\rho\hat f(0)+\frac{b}{c^3}+i\frac{\rho\hat f^2(0)\omega}
{4\pi c^3}\right)^{-1}
\end{equation}
and from eq.~(\ref{dos}) one gets exactly the Debye's law,
$g(\omega)=\omega^2/(2\pi^2\rho c^3)$, in the $\omega\to 0$ limit.  As
a first step in generalizing this result, we note that the same
conclusion is obtained by considering the limit of high sound velocity
(i.e.\ $1/c^3$ small with respect to $\rho\hat f(0)$).

By decreasing the density we easily show that in eq.~(\ref{ginf}), due
to the square root term, $G^\infty(0)$ develops an imaginary part when
$\rho<\rho_c$, where $\rho_c$ is fixed by $(b/c^3-\rho_c\hat
f(0))^2=4\rho_c a/c^3$.  We claim that this can be interpreted as a
phase transition in the space of the eigenvalues of the Hessian matrix
driven by the value of the density. The two phases separated from such
a phase transition are the stable phase (all positive eigenvalues) and
the unstable phase (negative and positive eigenvalues).  The order
parameter is $\varphi = -\textrm{Im}\,G^\infty(i0^+)$ which vanishes
as $\varphi\sim |\Delta|^\beta$, with $\beta=1/2$. The value of the
exponent of course could depend on the cactus approximation and is
highly remindful of mean field theories. Most interestingly, the Boson
peak arises, in the stable phase, as a signature of the phase
transition. In fact, by setting $\Delta\equiv (\rho-\rho_c)$ and
expanding for small $\Delta$,
\begin{equation}
-\textrm{Im}\ G^\infty(\omega^2+i0^+)\propto
\textrm{Im}\sqrt{\alpha\Delta -i\omega/\omega^*} \quad \Longrightarrow \quad
\left\{
\begin{array}{lr}
g(\omega)\sim \omega^{3/2} &
\,\, \textrm{if}\,\, \omega/\omega^*\gg\alpha\Delta\\
g(\omega) \propto \omega^2/\sqrt{\alpha\Delta} &
\,\, \textrm{if}\,\, \omega/\omega^*\ll\alpha\Delta
\end{array}
\right. ,
\label{Im}
\end{equation}
where $\alpha$ is a positive constant and $\omega^*\equiv
2\pi\sqrt{a/{\rho_c c^3}}$. Hence there exists a frequency that
signals a crossover from a Debye behaviour to a different kind, namely
$g(\omega)\sim\omega^\gamma$, $\gamma=3/2$.  We identify the Boson
peak with such frequency, i.e. $\omega_{BP} = \alpha\Delta\omega^*$.

From the experimental point of view, this implies that the BP is
indicated from a peak in the function $g(\omega)/\omega^2$m not in
$g(\omega)$. The BP frequency moves toward $0$ when approaching the
transition (from the stable side) and its height diverges.
Eq.~(\ref{s-dos}) shows that at the level of the one-phonon
approximation it can also be detected in the $p \to \infty$ limit of
the dynamic structure factor $S(p,\omega)$.

As said above, the control parameter need not be the density. For
example, by fixing the density and letting the sound speed to vary, we
find that
\begin{equation}
\varphi\propto\textrm{Im} \sqrt{(b-\rho c^3\hat f(0))^2-4\rho ac^3}. 
\end{equation}
Then the critical sound speed is fixed by the condition $(b-\rho
c_c^3\hat f(0))^2=4\rho ac_c^3$. Now $\Delta = c-c_c$ and once again
$\varphi=\alpha'|\Delta|^\beta$ with $\beta =1/2$.

Since the quantities $A$, $b$ and $\hat f(0)$ depend on the
thermodynamic parameters, in principle each of them could be chosen as
a control parameter to describe this phase transition. However, one
parameter (the potential energy) seems to be more physically
meaningful than the others when the energy landscape approach to the
glass transition is taken into account. It has been shown indeed in
numerical simulations that the typical stationary point of the
Hamiltonian closest to equilibrium configurations is a saddle above
the mode-coupling temperature and a minimum below
\cite{saddles}. Hence eq.~(\ref{asint}) describes (in an approximate
way) this phase transition in the space of all stationary points
(saddle and minima). This we shall call the {\em saddle-phonon
transition}.

Since the number of eigenvalues of a stationary point depends only on
its energy (in the thermodynamic limit)~\cite{saddles}, it plays a
special role as a control parameter in the study of the Boson
peak. Stationary points (and their energy), though, are easily
accessible in simulations but not in experiments. It is thus better to
formulate the scaling laws arising from the theory without any
reference to the parameter chosen. We then recast eq.~(\ref{Im}) as
\begin{equation}
g(\omega,\Delta)=\omega^\gamma h(\omega\Delta^{-\rho}), 
\qquad h(x)\sim \left\{
\begin{array}{lcr}
x^{2-\gamma}&&x\ll 1\\
\textrm{const.}&&x\gg 1
\end{array}
\right. ,
\end{equation}
with $\Delta$ defined in terms of an arbitrary control parameter.
Since the Boson Peak signals a crossover between the two phases, this
scaling law implies that $\omega_{BP}\sim \Delta^\rho$ and
$g(\omega_{BP},\Delta)/\omega_{BP}^2\sim\Delta^{-\eta}$, with
\begin{equation}
\eta=\rho(2-\gamma)\; .
\end{equation}
Since the result is model-independent this law is expected to hold
very generally, and in the next section we will check that numerically
in a simplified model. The ERMT (in the cactus approximation) predicts
$\rho = 1$, $\gamma = 3/2$, $\eta=1/2$. Numerical results on a fragile
glass former have already turned out to be in a reasonable agreement
with such scaling laws~\cite{nature}.

A similar asymptotic analysis gives the scaling behaviour of the
linewidth $\Gamma(p)$ of Brillouin peaks. Interestingly enough, it
changes at the saddle-phonon transition. From the analytic point of
view, the longitudinal and transverse $\Gamma(p)$ are given by the
corresponding imaginary part of the self-energy computed at the
position $\omega_0$ of the peak. In both cases the leading
contribution in the limit of small momenta is obtained by considering
the large $q$ contribution in the integral of eq.~(\ref{eqint}):
\begin{equation}
\textrm{Im}\,\Sigma({\bf p},\omega)\simeq -\frac{\pi g(\omega)}
{2\omega}\int\frac{d^3 q}{(2\pi)^3} V^2({\bf q},{\bf p}) .
\end{equation}
In the stable phase the Debye regime ($g(\omega)\sim\omega^2$) at low
enough frequencies ($\omega < \omega_{BP}$) implies
\begin{equation}
\Gamma_{L,T}(p) \sim
g(\omega) p^2/\omega^2 \sim p^2 .
\end{equation}
This is the asymptotic result both in the longitudinal and the
transversal case, but in practice one should be very careful when
fitting the experimental results with the above law. In fact, when
approaching the transition from the stable phase, the Boson peak
frequency $\omega_{BP}$ shifts to zero and the Brillouin peak might
fall in the region where $g(\omega) \propto \omega^{3/2}$ yielding a
flattening of the scaling law (the exponent would be $3/2$ instead of
$2$). Moreover, a crossover from $p^2$ to a $p^4$ behaviour in the
region where the dispersion relation is still linear cannot be ruled
out. In general, both the $p^2$ and $p^4$ terms are present, and their
relative weights depend on the thermodynamic parameters
\cite{erm2,erm2bis}.

In the unstable phase instead one has $g(\omega)\sim \omega$, due to
the fact that $\mbox{Im}\, G^\infty(0) \neq 0$ and to eq.~(\ref{dos}),
then the broadening becomes
\begin{equation}
\Gamma_{L,T}\sim p .
\end{equation}
This theoretical result is very suggestive because the possibility
arises of investigating the saddle-phonon transition through
measurements of the Brillouin peak.

We stress that the results of this section are model-independent.
Next section we will test them in the simplest case, the Gaussian
model.

\section{Numerical results for a Gaussian model}

Here we solve numerically the cactus equation for the case where
$f(p)$ has a Gaussian form and compare with direct numerical results
for the same model. This will confirm that the saddle-phononn
transition described by the Euclidean Random Matrix theory is not an
artifact of the approximation involved (cactus resummation).  The
model is described by
\begin{equation}
\hat f_{\mu\nu}({\bf p}) = \hat f_L(p)\frac{p_\mu p_\nu}{p^2} +
\hat f_T(p) \left(\delta_{\mu\nu}-\frac{p_\mu p_\nu}{p^2}\right),
\qquad f_{L,T}(p)=\left(\frac{2\pi}{\sigma_0^2}\right)^{3/2}
\exp{(-p^2/2\sigma^2_{L,T})} .
\label{model}
\end{equation}

\begin{center}
\begin{figure}
\psfrag{Fgrviir}[][][1.1]{$F[g(r)\partial_{\mu\nu} v(r)]$}
\psfrag{w/w0}[][][1.1]{$\omega^0_{L,T}(p)$}
\psfrag{pipo}[][][1.1]{$p/p_0$}
\includegraphics[width=0.6\columnwidth]{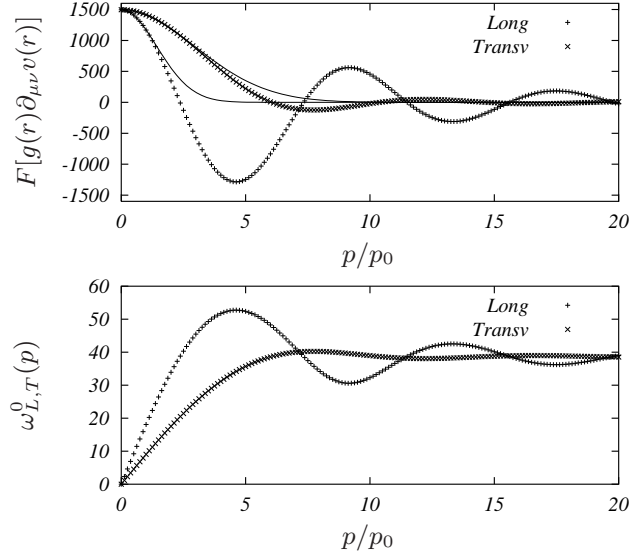}
\caption{\small Top: The Fourier transform (longitudinal and
transverse components) of $g(r)\partial_{\mu\nu} v(r)$ ($g(r)$
is the radial distribution function and $\partial_{\mu\nu}
v(r)$ is the second derivative of the pair potential) as obtained from
a simulation of a soft sphere system at $T=16$. The Gaussian fits
(also shown as solid lines) are made in the small momentum region, and
the ratio of two variances (which is in our model the square of the
ratio of the longitudinal and transverse sound speed) is $\simeq
2$. This is the value we fixed in the model. Bottom: The $bare$
dispersion relations (see text) $\omega^0_{L,T}(p) = \rho[\hat
f(0)-\hat f(p)]^{1/2}$ corresponding to this system. They saturate at
the same value but the asymptotic behaviour is reached for values of
the momenta greater than the ones studied in experiments.}
\label{ff}
\end{figure}
\end{center}

\begin{center}
\begin{figure}
\psfrag{wiwo}[][][1.2]{$\omega/\omega_0$}
\psfrag{wiwi}[][][1.2]{$\omega/\omega^{'}_0$}
\psfrag{gw}[][][1.2]{$g(\omega)$}
\includegraphics[width=0.75\columnwidth]{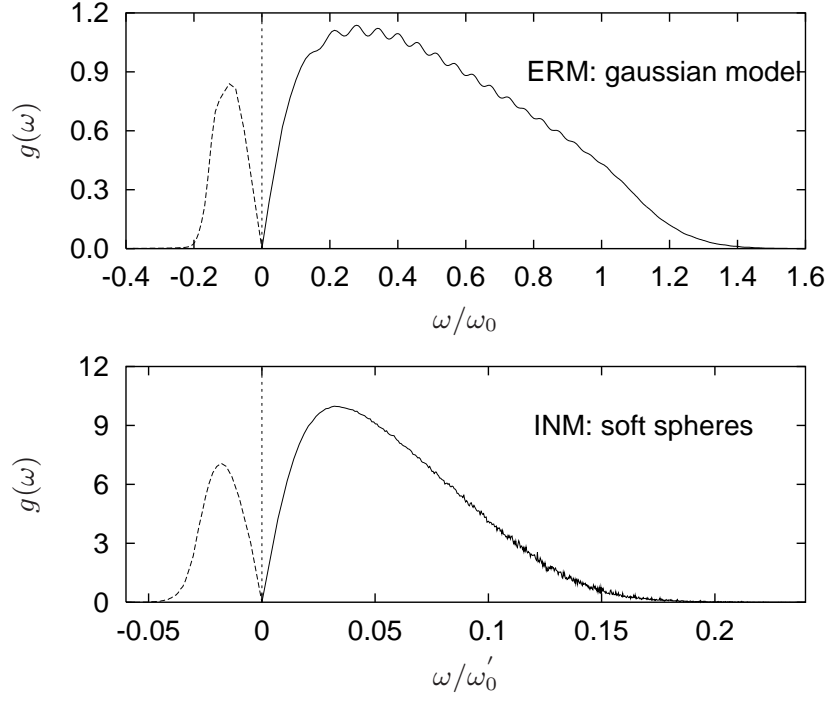}
\caption{\small Top: The density of states for our ERM model in the
unstable phase ($\rho = 0.2$), as obtained numerically from the method
of moments. This numerical method allows to build the density of
states of a large matrix (here the Hessian of a $N=10^6$ particle
system) by a sum of delta-functions. The width of these is responsible
for the oscillations in the resulting VDOS that should vanish once one
averages over a big number of samples. $\omega_0$ is the saturation
frequency for high values of momenta in the dispersion
relation. Bottom: The INM spectrum as arising from a simulation of a
soft sphere system of $N=2048$ particles at $T = 0.68$, i.e. in the
liquid phase. We average here over $O(10^2)$ samples, so there are not
big oscillations in the VDOS. As before, $\omega_0^{'}$ is the frequency
corresponding to the saturation of the dispersion relation. In both
cases the density of states for imaginary frequencies is plotted in
the negative real axis ($i\omega\to -\omega$), as usual.}
\label{cfr}
\end{figure}
\end{center}

This choice for $\hat f(p)$ is mainly due to its simplicity.  However,
we have shown that many features, ranging from the behaviour of the
Boson peak close to the saddle-phonon transition to the width of the
Brillouin peak in the stable and unstable phases are independent of
the details of the model. Hence we expect the Gaussian model to be as
good as any other to study such features.  Moreover, as we discussed
at the end of sec.~II, the superposition approximation amounts to
taking $\hat f_{\mu\nu}(p)={\cal F}[g(r) v_{\mu\nu}(r)]$, meaning that
$\hat f_{L,T}(p)$ are finite at $p=0$. Thus a Gaussian approximation
for $\hat f(p)$ at low enough momenta is always possible. As an
example, in Fig.~\ref{ff} (top) we show the longitudinal and
transverse components of the Fourier transform of $g(r)v_{\mu\nu}(r)$
for a soft sphere system (i.e.\ a one-component liquid with pair
potential $v(r)=r^{-12}$) and the comparison with a Gaussian fit. The
fit gives $\sigma_T \simeq 2\sigma_L$ (hereafter we shall fix
$\sigma_L$ to $\sigma_0$, the lenght scale of the problem). We can
check that the Gaussian model is able to reproduce the qualitative
behaviour of a real system by looking at Fig.~\ref{cfr}, where it is
shown that the density of states of a Gaussian euclidean random matrix
looks qualitatively like the INM spectrum of a liquid. Thus a Gaussian
model is not an outrageous approximation.

We shall consider various values of the density, which is here a
possible control parameter, comparing the analytical (cactus) results
with the numerical spectra and dynamic structure factor obtained from
the method of moments~\cite{mom}. This method allows one to obtain the
density of states and the dynamic structure factor of a given $N\times
N$ matrix up to $N=O(10^6)$ in a reasonable CPU time.  The numerical
solution of eq.~(\ref{eqint}), which is actually two (coupled)
equations for the longitudinal and transverse parts of the resolvent,
requires some work.  It can be solved by iteration at fixed $z$ as
long as one can do the three-dimensional integral sufficiently
fast. To do so, we write the equations as convolutions by expanding
the vertex $V({\bf q},{\bf p})$: this gives four terms, which can be
evaluated with two convolutions. These can be calculated very
efficiently using the fast Fourier transform (FFT), as long as the
functions are sufficiently smooth. Since this is not the case, the
resolvent must be separated in a regular part which tends to zero at
infinity, and a part which is quasi-singular and has a finite
limit. After the convolution of this part is worked out analytically,
the remaining convolutions can be done using FFT. In this way, each
iteration takes a time $O(N\log N)$, and an iterative scheme can be
reasonably implemented.

\subsection{$\rho \gg \rho_c$: the stable phase (phonons)}

\begin{figure}
\centerline{\psfig{figure=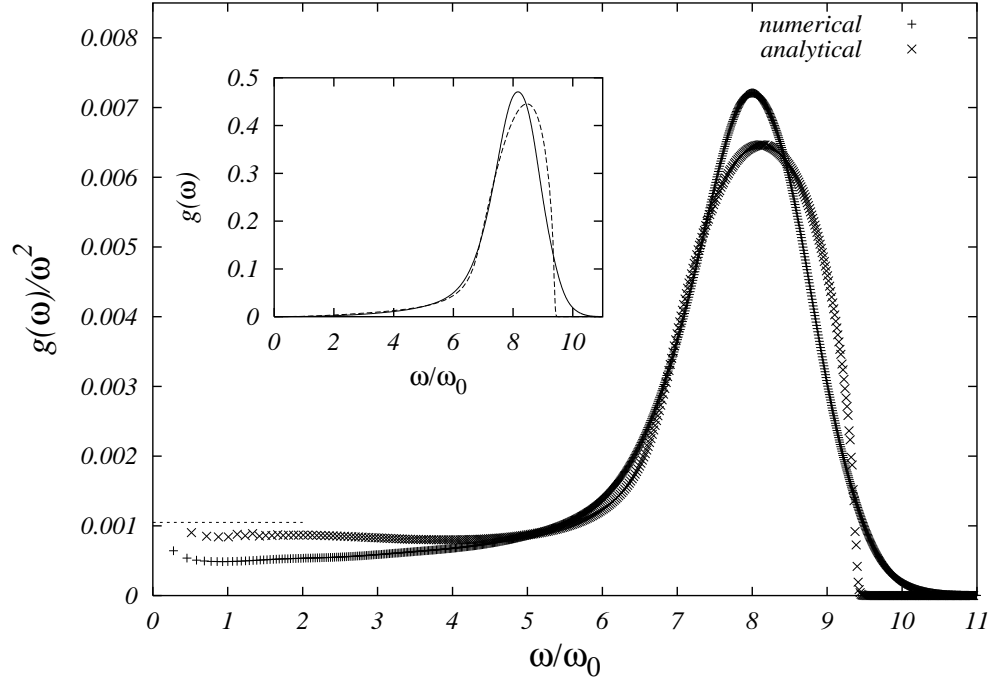,width=0.75\columnwidth,angle=0}}
\caption{\small The VDOS g($\omega$) as a function of eigenfrequencies
divided by the predicted Debye behaviour $\omega^2$ for
$\rho=4>\rho_c$. In the inset, we show g($\omega$) vs. $\omega$ both
numerical (obtained via the method of moments) and analytical (from
the numerical solution of the cactus equation~(\ref{eqint})).}
\label{giom}
\end{figure}

In the high density regime we are in the phonon or stable phase, since
all the eigenvalues are positive. Fig.~\ref{giom} shows the VDOS, and
its Debye behaviour ($\propto \omega^2$) for small frequencies.  In
this regime the approximations used in deriving the integral
equation~(\ref{asint}) are quite good since the analytic solution
reproduces the numerical spectrum (and in particular the Debye
behaviour) rather accurately. The large peak at high frequency arises
from a pileup of states due to the flattening of the dispersion
relations. Those excitations are the non-propagating glassons
described by the high-frequency limit of eq.~(\ref{asint}). In a way,
the peak can be considered as the off-lattice analogous of the van
Hove singularity, the main difference being that density fluctuations
smooth the cusp, giving it rather a semicircular form.  Finally, let
us note that the analytical solution misses the high-frequency tail of
the VDOS. These modes are expected to appear in a number which is
exponential in $\rho$, and thus cannot be recovered in a $1/\rho$
expansion.

\begin{figure}
\centerline{\psfig{figure=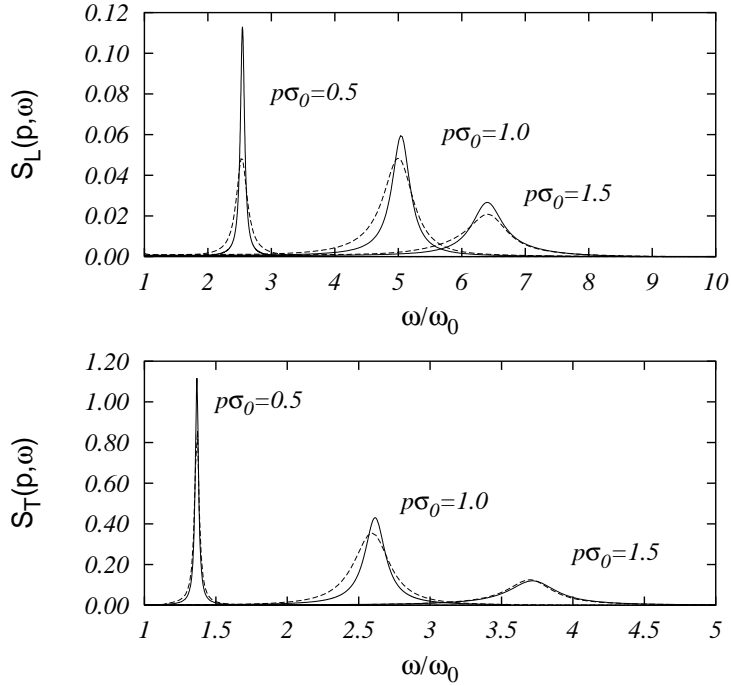,width=0.75\columnwidth,angle=0}}
\caption{\small The longitudinal (top) and transverse dynamic
structure factor in the high density (stable) region for three values
of exchanged momenta. The comparison between analytical (solid
line) and numerical (dashed line) data shows that there is some
difference especially for small values of momenta. The range of
frequencies is in the Debye regime (see Fig.~\ref{giom}).}
\label{SLST4}
\end{figure}

In Fig.~\ref{SLST4} we plot the longitudinal and transverse components
of the dynamic structure factor for three values of $p$. The agreement
between numerical and analytical solutions is better at the higher
values of $p$, though the position of the Brillouin peak is correctly
reproduced at all $p$. Since the peaks are well defined, we are
allowed to study the dispersion relations as well as the linewidth
$\Gamma$ as a function of $p$. They are obtained by fitting the peaks
with $S^{(1)}(p,\omega) \propto \omega_{L,T}^2(p)\Gamma_{L,T}(p) /
\left[ (\omega^2-\omega_{L,T}^2(p))^2+\omega^2\Gamma_{L,T}^2(p)
\right]$, following the experimental procedure~\cite{exp}.  While the
agreement about the numeric and analytic dispersion relations is
striking, there remains some discrepancy in the linewidth, especially
at the lower momenta of the longitudinal case. It is worthwhile to
note that both dispersion relations saturate at the same value at vary
large momenta, though they can be very different in the regime which
is presently explored by simulations and experiments ($p \sigma_0 \sim
1$). Interestingly enough, both relations coincide with the bare
dispersion relation $\omega_0(p)$ of eq.(\ref{nuda}), the
renormalization due to the self-energy in (\ref{allargamento}) seeming
negligible. As for the linewidth, the numeric and analytic (integral
equation) results are in reasonable agreement with the asymptotic
theoretical prediction $\Gamma\sim p^2$.  Furthermore, the scaling
seems to be independent from the longitudinal or transverse nature of
the excitations. However, it cannot be ruled out from our data the
possibility that a crossover between $p^2$ and $p^4$ is verified
before saturating at the limiting value.

\subsection{$\rho\simeq\rho_c$: near the transition. }

\begin{figure}
\centerline{\psfig{figure=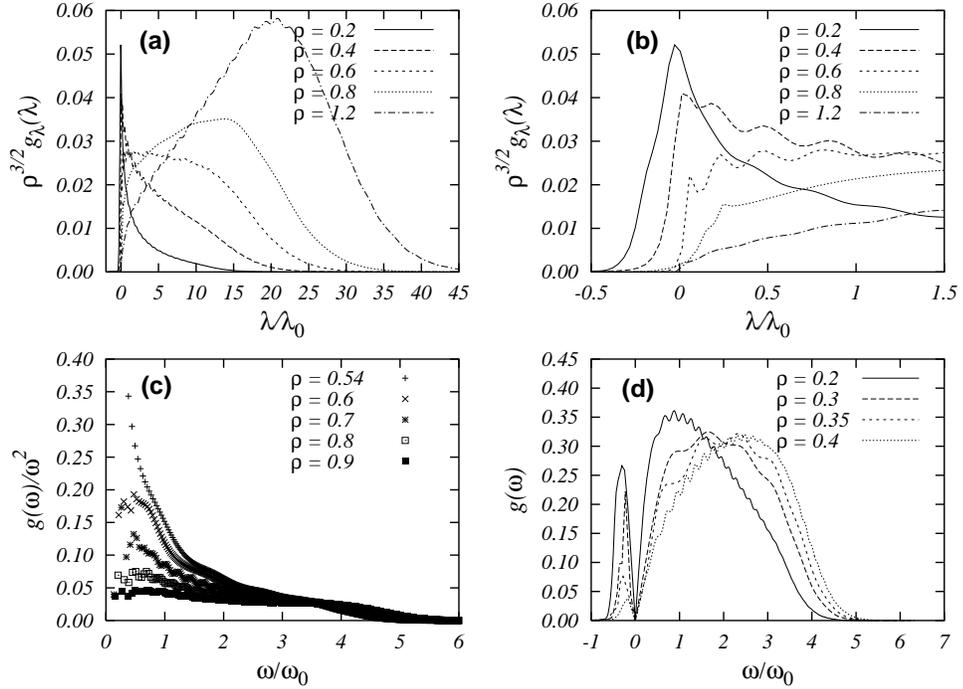,width=0.75\columnwidth,angle=0}}
\caption{\small (a): The spectrum of the Gaussian ERM as obtained
numerically with method of moments for several values of the density
$\rho$, below and above the critical density $\rho_c\simeq 0.54$. All
values of the densities are given in units of $\sigma_0^{-3}$ (see
eq.~(\ref{model})). (b): Same data, a zoom of the interesting region
near $\lambda =0$. The value of $g(0)$ grows on lowering the
density. (c): Plot of $g(\omega)/\omega^2$ vs.\ $\omega$ for densities
$\rho\gtrsim\rho_c$. The Boson peak is defined as a peak in this plot;
we see that a BP appears on lowering the density. (d): The VDOS as a
function of real (positive axis) and imaginary frequencies (negative
axis).}
\label{num}
\end{figure}

In Fig.~\ref{num} we show the spectrum of Gaussian matrices (in terms
of eigenvalues as well as frequencies) as obtained from the method of
moments for several values of the density near the critical one, which
turns out to be $\rho_c\simeq 0.54$. If we lower the density we enter
the unstable region and this is revealed by the appearance of an
extensive fraction of unstable modes (imaginary frequencies). A BP
appears near the transition point in the low frequency region.

\begin{figure}
\centerline{\psfig{figure=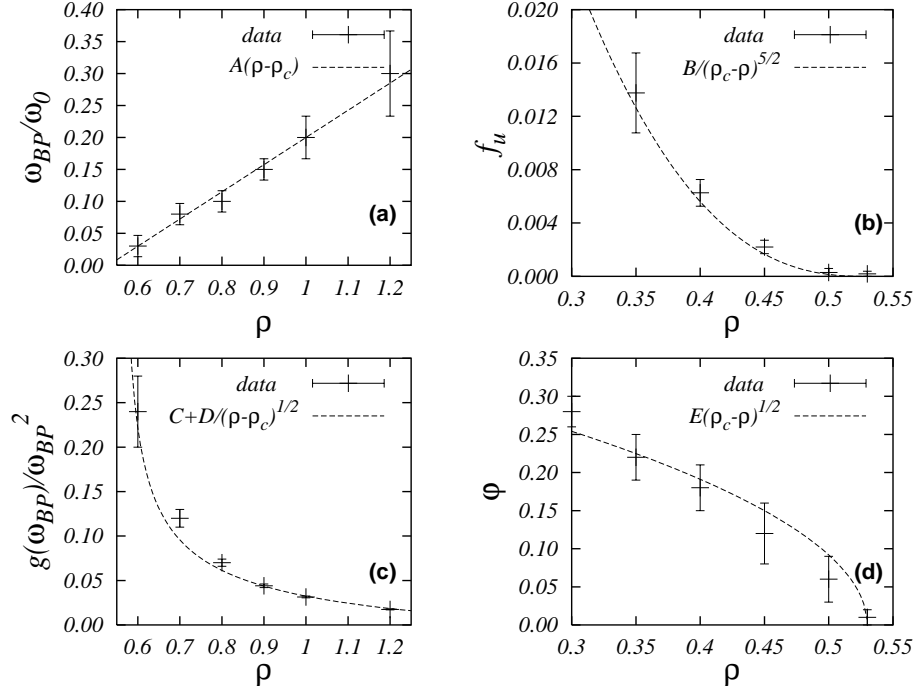,width=0.75\columnwidth,angle=0}}
\caption{\small Data taken from numerics (method of moments). The
critical density in the following fits has been fixed to $\rho_c=0.54$
and capital letters are the fitting parameters. (a): The position of
Boson peak is plotted as a function of the density near the critical
point. We find that $\omega_{BP}$ vanishes linearly in
$\Delta=\rho-\rho_c$. (b): The fraction of unstable modes vanishes as
$(\rho_c-\rho)^{2.5}$, thus giving a value $\gamma=3/2$. (c): The
height of the BP, defined by $g(\omega_{BP})/\omega_{BP}^2$, diverges
as $\Delta^{-\eta}$, with $\eta=1/2$. (d): The order parameter
$\varphi\equiv -\mbox{Im}\, G^{\infty}(0)$ (see text) vanishes as
$(\rho_c-\rho)^\beta$, with $\beta = 1/2$.}
\label{bp_num}
\end{figure}

We now obtain the exponents of the transition (Fig.~\ref{bp_num}).
Figs.~\ref{bp_num}a and \ref{bp_num}c show that the position of the BP
is linear with respect to $\Delta\equiv (\rho-\rho_c)$ and that the
height of BP diverges as $\Delta^{-1/2}$. This confirms the
theoretical predictions $\nu=1$ and $\eta=1/2$. In Fig.~\ref{bp_num}b
we determine the value of $\gamma$ by studying the fraction of
unstable modes.  Indeed, from eq.~(\ref{Im}), in the region of
parameters such that $\rho\lesssim \rho_c$, we can argue that the VDOS
behaves as $g(\omega)=\omega^{\gamma}\tilde g(\omega/|\Delta|)$, where
$\tilde g(x)$ is a scaling function and $\gamma=3/2$. The fraction of
unstable modes is defined as $f_u=\int_{-\infty}^0
g_\lambda(\lambda)d\lambda$. We thus have that
\begin{equation}
f_u(\Delta)=\int_0^\infty\!\! d\omega\, \omega^\gamma \tilde
g(\omega/|\Delta|) \sim|\Delta|^{1+\gamma} .
\end{equation}
We find $f_u\sim(\rho_c-\rho)^{5/2}$, i.e.\ $\gamma=3/2$. Finally,
the order parameter $\varphi$ vanishes as $(\rho_c-\rho)^\beta$ with
$\beta=1/2$ (Fig.~\ref{bp_num}d).

\begin{figure}
\centerline{\psfig{figure=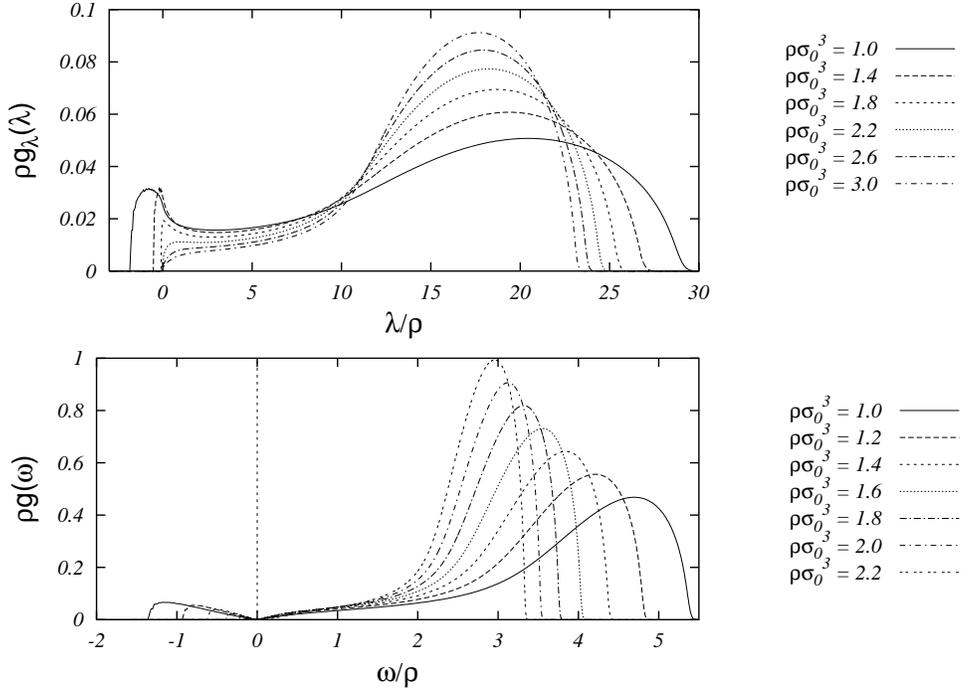,width=0.75\columnwidth,angle=0}}
\caption{\small Top: The density of eigenvalues obtained from the
numerical solution of equation (\ref{eqint}) at fixed $c_L/c_T=2$ for
several values of the density $\rho$. The transition from a stable
phase to an unstable phase is clear since an extensive number of
negative eigenvalues appears. Bottom: The transition can be seen also
by plotting the spectra as a function of the frequencies, with the
unstable modes (i.e. imaginary frequencies) on the negative real
axis.}
\label{rho}
\end{figure}

\begin{figure}
\centerline{\psfig{figure=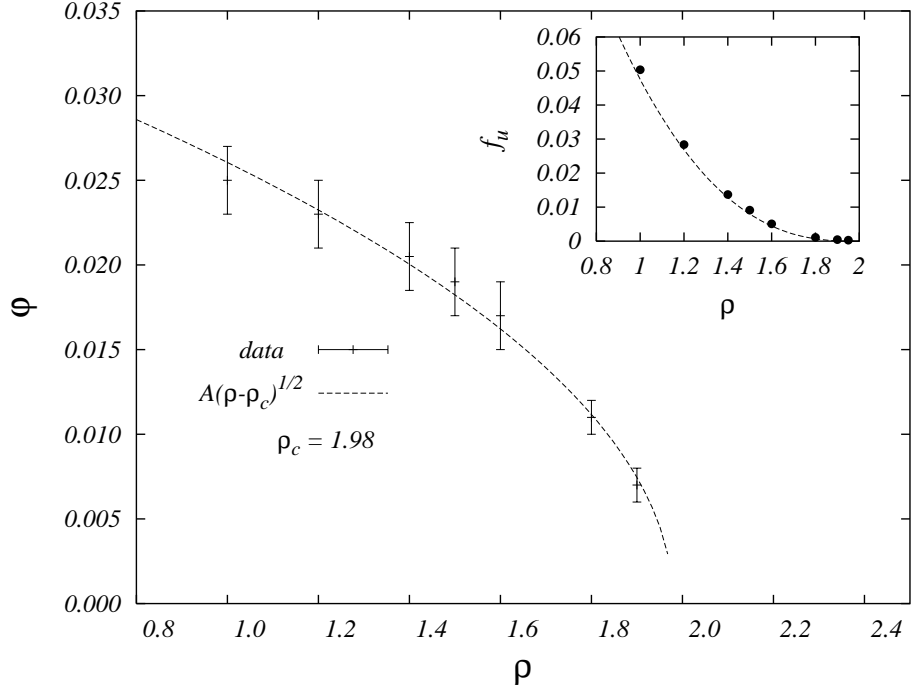,width=0.75\columnwidth,angle=0}}
\caption{\small The order parameter $\varphi\equiv$ -\mbox{Im}\,
G$^\infty$(E=0) (here obtained from the numerical solution of the
self-consistent equation~(\ref{eqint})) behaves as
$(\rho-\rho_c)^\beta$, with $\beta=1/2$, as predicted by the
asymptotic analysis of the equation. In the inset, we show that the
fraction of unstable modes vanishes at the transition as a power law
with the exponent $\gamma+1=5/2$.}
\label{rhoc}
\end{figure}

The phase transition is also found by the numerical solution of the
integral equation (eq.~\ref{eqint}), as one can see from
Fig.~\ref{rho}.  Actually, this solution does not give the spectrum
very accurately (one can still discern the semicircle centered in
$\omega\sim \rho \hat f(0)$ rather separated from the rest of the
spectrum, while we have seen that they mix at low enough density---
Fig.~\ref{num}) and it overestimates the critical density,
$\rho_c^\mathrm{analyt}\simeq 1.98 > \rho_c^\mathrm{numeric}$.  In
Fig.~\ref{rhoc} we check that the order parameter vanishes with the
correct exponent, $\beta=1/2$, and also for the fraction of unstable
modes. It is interesting to remark that these exponents are also found
for a larger class of nearly Gaussian matrices, corresponding to
functions $f(p)\propto \exp{\left(-p^2(1+a p^{2n})/2\sigma^2\right)}$.

\subsection{$\rho<\rho_c$: the unstable phase (saddles). }

\begin{figure}
\centerline{\psfig{figure=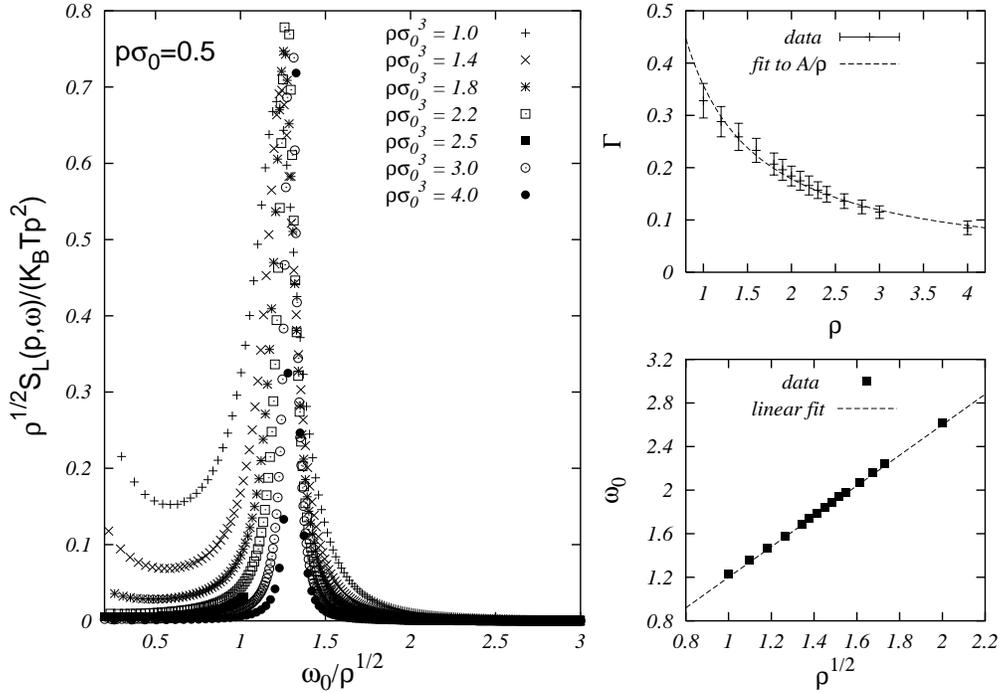,width=0.75\columnwidth,angle=0}}
\caption{\small The position of the Brillouin peak $\omega_0$, here
obtained from the numerical solution of the integral
equation~\ref{eqint}, scales as $\sqrt{\rho}$ in a wide range of
densities across the critical one. The linewidth $\Gamma$ decreases as
$1/\rho$ in the same range of densities.}
\label{brip5}
\end{figure}

\begin{center}
\begin{figure}
\psfrag{GL}[][][1.4]{$\Gamma_L$}
\psfrag{ps0}[][][1.4]{$p\sigma_0$}
\psfrag{pi}[][][1.2]{$p$}
\psfrag{wiwo}[][][1.4]{$\omega/\omega_0$}
\psfrag{wL}[][][1.4]{$\omega_L$}
\psfrag{p2SLpw}[][][1.4]{$p^2S_L(p,\omega)$}
\psfrag{woLp}[][][1.2]{$\omega^0_L(p)$}
\psfrag{ps0=025}[][][1.2]{$p\sigma_0 = 0.25$}
\psfrag{ps0=05}[][][1.2]{$p\sigma_0 = 0.5\;\;\;$}
\psfrag{ps0=075}[][][1.2]{$p\sigma_0 = 0.75$}
\psfrag{ps0=10}[][][1.2]{$p\sigma_0 = 1.0\;\;\;$}
\includegraphics[width=0.75\columnwidth]{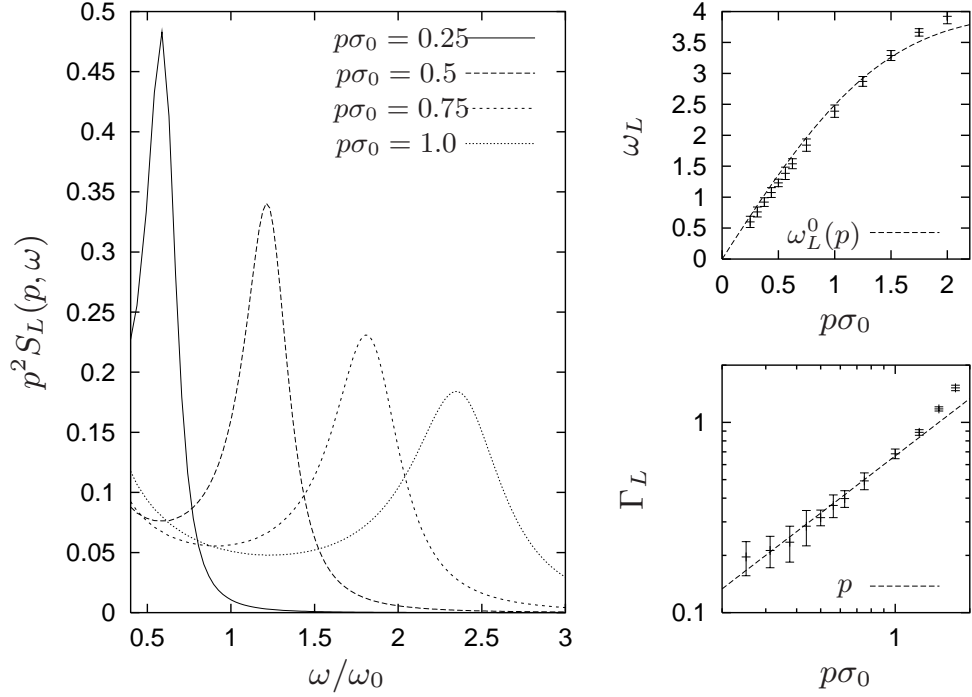}
\caption{\small The structure factor in the unstable phase, from the
solution of eq.~(\ref{eqint}), shows Brillouin peaks (left); the
dispersion relation (top right) obtained from these peaks is similar
to that in the stable phase. The linewidht scaling, however, is
changed to $\Gamma \sim p$ (bottom right).}
\label{brill-unst}
\end{figure}
\end{center}

From the results of the integral equations, we show in
Figs.~\ref{brip5} and~\ref{brill-unst} that the Brillouin peaks still
exist in the unstable phase (below $\rho_c$).  The main difference
with the stable phase is the peak around $\omega \sim 0$, due to the
negative eigenvalues. Of course, since ERMT is a purely harmonic
theory, it is expected to describe correctly only frequencies much
greater than the inverse of the $\alpha$ (structural) relaxation time
$\tau_\alpha$. So as long as $\omega \tau_\alpha \gg 1$, our
computation should be reliable to study also the saddle phase. In
fact, our finding of Brillouin peaks even when negative eigenvalues
are present is consistent with the fact that in real systems Brillouin
peaks related to the high-frequency sound are found even well above
the mode-coupling temperature \cite{exp}, which corresponds to the
saddle-phonon transition.

Furthermore it is worthwhile to notice that, as predicted from the
analytic asymptotic analysis, the saddle-phonon transition produces a
change in the linewidth scaling.  In Fig.~\ref{brill-unst} we show
that from the integral equation one gets $\Gamma(p) \sim p$. It would
be very important to confirm this result in simulations of realistic
models or, better, in experiments.  Finally, at fixed momentum, the
scaling laws $\omega_0\sim \rho^{1/2}$ for the position of the peak,
and $\Gamma \sim \rho^{-1}$ for its width remain unchanged across the
transition (Fig.~\ref{brip5}).

\subsection{The sound velocity as a control parameter}

\begin{figure}
\centerline{\psfig{figure=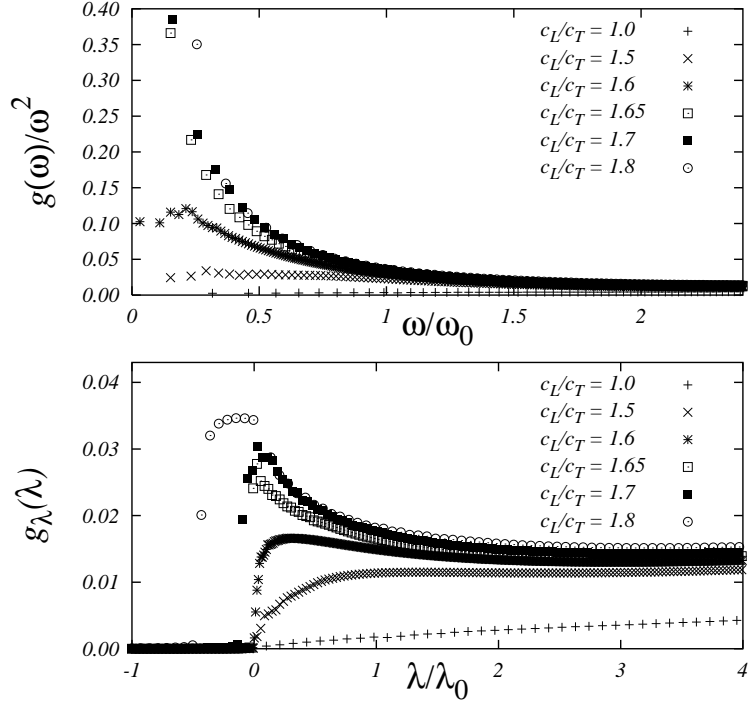,width=0.75\columnwidth,angle=270}}
\caption{\small Data obtained from the solution of eq.~\ref{eqint}. At
fixed density $\rho=1$ , the VDOS (evaluated from the integral
equation) depends on the ratio between longitudinal and transverse
sound velocity in the low energy region. This is exactly the same as
varying the value of the sound speed, since
$c\propto(1+2(c_L/c_T)^{-3})^{-1/3}$.}
\label{sT}
\end{figure}

\begin{figure}
\centerline{\psfig{figure=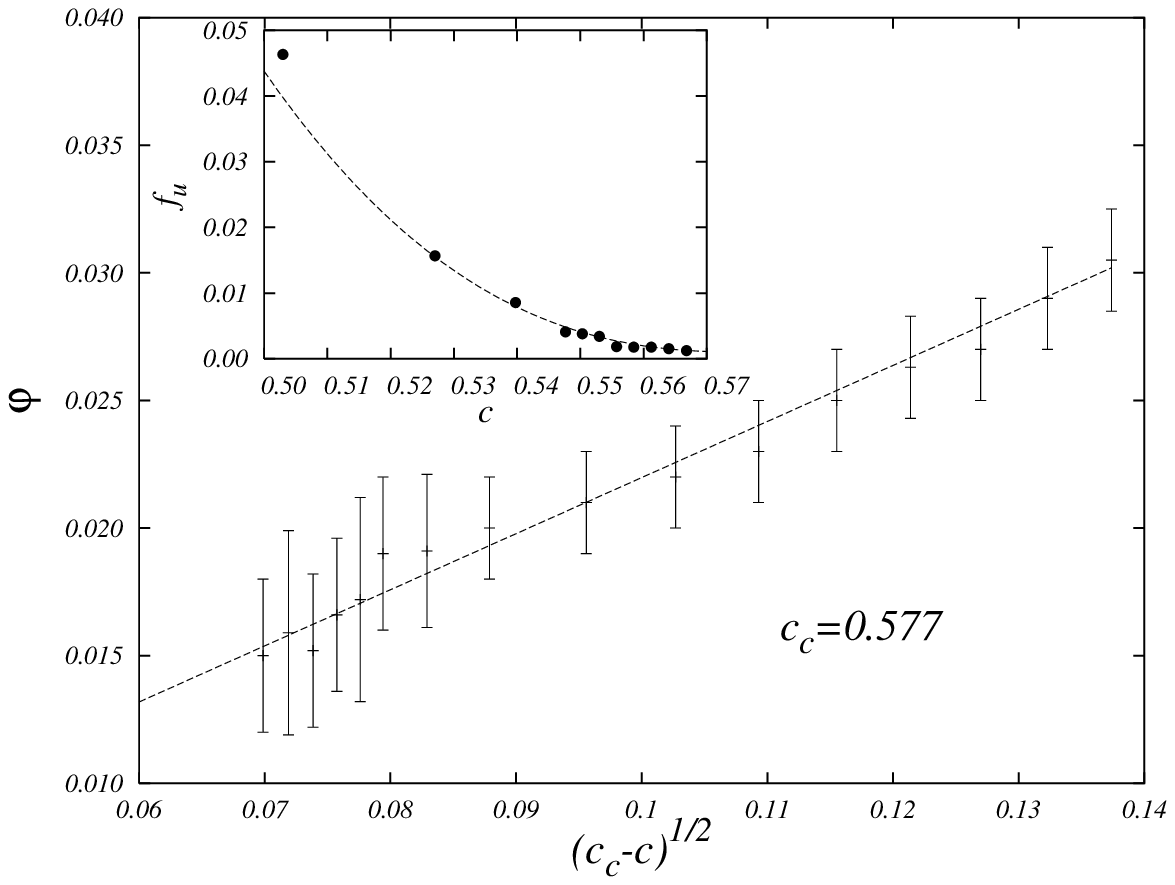,width=0.75\columnwidth,angle=0}}
\caption{\small At fixed density ($\rho=1$ here) the transition is
driven by the speed of sound, and the order parameter vanishes as a
square root. In the inset we show that the fraction of unstable modes
vanishes as $(c-c_c)^{2.5}$, as predicted by ERMT. The data are
obtained from the numerical solution of the cactus equation.}
\label{cc}
\end{figure}

Finally, we look at the instability transition as driven by the speed
of sound, as discussed at the end of section~III. Since
$c\propto(1+2(c_L/c_T)^{-3})^{-1/3}$, this is determined by the ratio
between longitudinal and transverse sound velocities, i.e. by the
anisotropy. We thus have studied our ensemble of Gaussian matrices at
a fixed density $\rho =1$ as a function of the ratio $c_L/c_T$. The
control parameter is now $\Delta =(c-c_c)$, and we find that
$c_c\simeq 0.58$. In Fig.~\ref{sT} we see that a BP appears on
approaching the critical value, which signals the transition to a
region with an extensive number of negative eigenvalues. Fig.~\ref{cc}
shows that the critical exponents are the same as when the density is
the control parameter. This confirms the universal character of the
phase transition discussed before.

\section{Conclusions}

In this work we have extended ERM theory to the physical case in which
both longitudinal and transverse vibrations are present. Even if the
model is exceedingly simple (harmonic vibrations around fully
disordered oscillations centers) it has many features in common with
real glasses and supercooled liquids:
\begin{enumerate}
\item A stable phase (no negative eigenvalues) is present, where
low energy excitations can be considered as phonons, in the sense that
they have a well defined wavenumber and that a Debye density of states
is found ($g(\omega)\propto\omega^2$).
\item There is a phase transition from the stable phase to a
saddle-phase, which in our model can be controlled by the
density. Numerical simulations~\cite{saddles} of supercooled liquids
have strongly suggested that such a phase transition occurs at the
mode-coupling temperature. In our model, the transition leaves a clear
mark in the low energy part of the spectrum in the stable (minima)
phase: an excess of states, that can be described with scaling laws
and critical exponents. We propose to identify this spectral feature
with the Boson peak of supercooled liquids and glasses. This implies
that our scaling laws should apply to experimental spectra.  We have
also found scaling laws for the approach from the unstable (liquid)
side of the transition, but it is still not clear how they could be
confronted with experiments.
\item In this stable phase, the scattering of sound waves is stronger
than in disordered crystalline systems: the linewidth scales like
$p^2$ {\em even in the harmonic approximation,} and there is a
significant mixing among longitudinal and transversal excitations. It
is interesting that, according to our calculation, the scattering of
sound waves is stronger in the unstable phase. To be precise, the
(harmonic) linewidth is found to be $\Gamma(p)\sim g(\omega)
p^2/\omega^2$, so that in a liquid phase, where $g(\omega)\sim\omega$,
one has $\Gamma(p)\propto p$ rather than $p^2$.
\end{enumerate}

We have studied our model using a high-density expansion, and we have
compared it with extensive numerical simulations. As one could expect,
the comparison is rather good at high density, in the phase where the
system is stable. On the numerical side, we have been able to study in
detail the vibrational spectrum in the neighborhood of the
phonon-saddle transition. Indeed, previous liquid
simulations~\cite{saddles,nature} had been carried out in very small
systems where low-frequency, long-wavelenght excitations were not
present, and our previously studied scalar model did not even have a
phonon-saddle transition. The present approximation overestimates the
critical value of the density by a factor four, a tremendous
improvement over the calculation for the scalar case, where an
inexistent transition was found due to the approximations. Since we
are dealing with a high density approximation it is not surprising to
find significant differences with the numerical calculations at low
densities. Nevertheless, universal features like scaling laws and
critical exponents are captured correctly, as the comparison with the
numerical simulation shows (surpringly, even the oversimplified scalar
model did capture the value of exponents). In any case, considering
transverse modes is a necessary step towards a quantitative theory of
glass vibrations. In fact, in systems with purely repulsive potentials
(like our Gaussian model, or soft spheres) it is vector displacements
that permit the rising of an instability, which would be otherwise
absent. For the Gaussian model, we have indeed verified that the
instability transition can also occur controlled by the ratio of the
sound velocities, and that the scaling laws apply as well with this
control parameter.

We think that, from the theoretical side, the subject is mature for a
detailed experimental study, since one has scaling laws and critical
exponents that will hopefully describe the experimental vibrational
spectra. Yet there are important experimental difficulties related
with the fact that the vibrational spectra could be reasonably
described as harmonic only at low temperatures. This is best explained
in terms of the potential energy landscape. A numerical simulation of
a supercooled liquid typically gets out of equilibrium at the
mode-coupling temperature. This means that the system is exploring the
energy minima close to the phonon-saddle transition where the Boson
peak should be a very prominent feature of the
spectrum. Unfortunately, in experiments one cannot cool as fast as in
a simulation, meaning that the system goes out of equilibrium rather
at the glass temperature (roughly speaking, the system goes too far
into the stable phase).  Instead, what we propose is to study the
harmonic spectrum (experiments done at low temperatures), of systems
that have fallen out of equilibrium at temperatures near the
mode-coupling temperature. We expect that this can be achieved by
using the new ultrafast cooling techniques
(hyperquenching). Preliminary results \cite{HYPERQUENCH} indicate that
indeed the Boson peak is enhanced at faster cooling, in qualitative
agreement with our expectations. However, much work is needed to check
to what extent the present results can quantitatively describe the
experimental spectra. In particular, the precise value of the critical
exponents could depend on the algebraic or exponential decay at large
distance of the interaction potential. This is not the case in the
present approximation.

Finally, an important result of the vectorial analysis is the fact
that the resolvent becomes isotropic in the infinite momentum
limit. This means that both dispersion relations saturate at the same
value (corresponding to the glasson region of the VDOS), and that both
structure factors tend to the VDOS for $p\to\infty$. This, strictly
speaking, precludes the identification of BP modes as those belonging
to the end of the transverse branch \cite{pilla02}. However, since the
dispersion relations can reach the limiting value in very different
ways, for intermediate values of $p$ (those presently studied in
simulations) they can be very different and may seem to saturate at
different values. As an example, in Fig.~\ref{ff} (bottom) we plot the
bare longitudinal and transverse dispersion relations of a monoatomic
soft-sphere liquid. Assume (as is the case for the Gaussian model)
that they are not significantly altered by higher order terms. Then we
see that the longitudinal branch shows larger fluctuations around the
limiting value than the transverse one and that they become similar
only at very large momenta. But at these valus of $p$ scattering
between the two branches is so large that the question of the
longitudinal or transversal nature of a given mode is rather
ill-posed. In any case, to better assess the role of the different
modes in more complex material, a model including optical branches is
needed. Thus we plan to investigate a binary model in the near future,
as well as to consider more realistic interactions within the same
formalism, along the lines discussed at the end of sect.~II.

\section*{Acknowledgment}
We are indebted with Giancarlo Ruocco, Oreste Pilla and Gabriele
Viliani for helpful discussions. V.M.-M. is a {\em Ram\'on y Cajal}
research fellow (MCyT, Spain). P.V. has been supported by the EC
network {\em DYGLAGEMEM} through a research fellowship.  This work has
been partly finantially supported by MCyT research contracts
FPA2001-1813 and FPA2000-0956.



\begin{figure}
\centerline{\psfig{figure=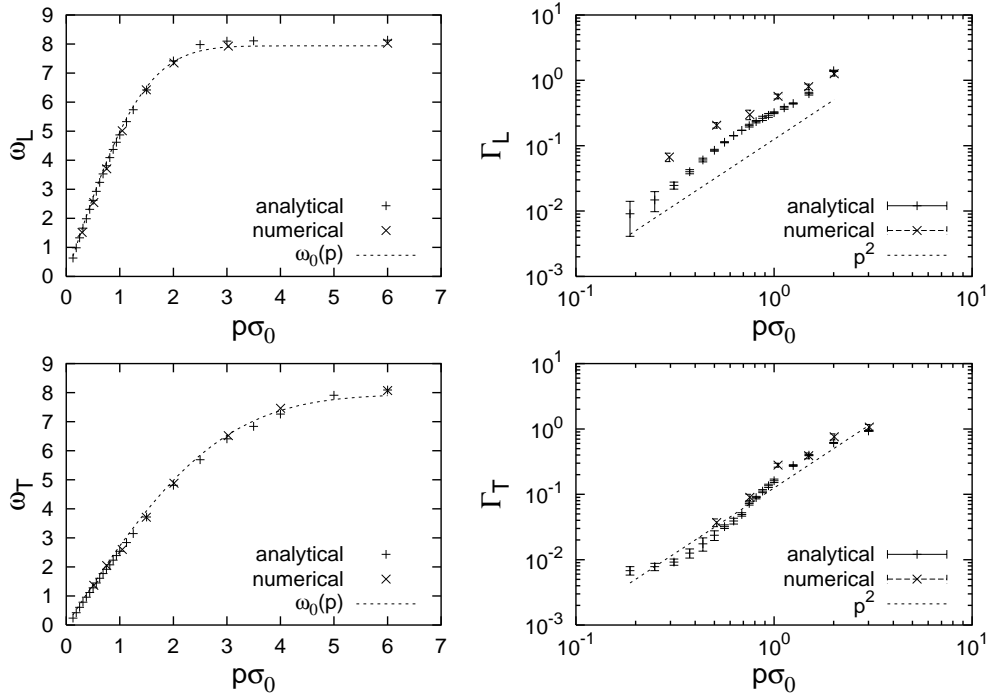,width=0.75\columnwidth,angle=0}}
\caption{\small The dispersion relations and the linewidth of the
Brillouin peaks for $\rho=4\gg\rho_c$, both from numerical evaluation
and from analytical computation (eq.~\ref{eqint}).}
\label{disp4}
\end{figure}

\begin{thebibliography}{11}

\bibitem{exp} F. Sette, M. Krisch, C. Masciovecchio, G. Ruocco, and
G. Monaco, Science {\bf 280,} 1550 (1998); G. Ruocco and F. Sette,
J. Phys.: Cond. Matt. {\bf 13}, 9141 (2001); E. Courtens , M. Foret,
B. Hehlen, R. Vacher, Solid State Commun. {\bf 117}, 187 (2001).

\bibitem{hansen} see for example J. P. Hansen and I. R. McDonald,
\emph{Theory of Simple Liquids}, Academic Press, London (1986).

\bibitem{ZHU91} T. C. Zhu, H. J. Maris, and J. Tauc , Phys. Rev. B {\bf
44,} 4281 (1991).

\bibitem{Ribeiro98} M. C. C. Ribeiro, M. Wilson, and P. A. Madden,
J. Chem. Phys. {\bf 108}, 9027 (1998).

\bibitem{scopigno03} T. Scopigno, S. N. Yannopoulos,
D. Th. Kastrissios, G. Monaco, E. Pontecorvo, G. Ruocco, and F. Sette,
J. Chem. Phys. {\bf 118}, 311 (2003).

\bibitem{debate} M. Foret, E. Courtens, R. Vacher, J.-B. Suck, Phys,
Rev. Lett. {\bf 77}, 3831 (1996); P. Benassi, M. Krisch,
C. Masciovecchio, V. Mazzacurati, G. Monaco, G. Ruocco, F. Sette, and
R. Verbeni, Phys, Rev. Lett. {\bf 77}, 3835 (1996); O. Pilla,
A. Cunsolo, A. Fontana, C. Masciovecchio, G. Monaco, M. Montagna,
G. Ruocco, T. Scopigno, and F. Sette, Phys, Rev. Lett. {\bf 85}, 2136
(2000); B. Ruffl\'e, M. Foret, E. Courtens, R. Vacher, and G. Monaco,
Phys. Rev. Lett. {\bf 90}, 095502 (2003).

\bibitem{elliott} see e.g. S. R. Elliott, \emph{Physics of Amorphous
Materials,} Longman, New York (1990) and references therein.

\bibitem{schirmacher98} W. Schirmacher, G. Diezemann, C. Ganter,
Phys. Rev. Lett. {\bf 81}, 136 (1998).

\bibitem{mayor00} V. Mart\'\i{}­n-Mayor, G. Parisi, and P. Verrocchio,
Phys. Rev. E, {\bf 62} 2373 (2000).

\bibitem{taraskin} S. N. Taraskin, Y. L. Loh, G. Natarajan, and
S. R. Elliott, Phys. Rev. Lett.  {\bf 86}, 1255 (2001); S. N. Taraskin
and S. R. Elliott, J. Phys.: Cond. Matt. {\bf 14}, 3143 (2002);
S. I. Simdyankin, S. N. Taraskin, M. Elenius, S. R. Elliott, and
M. Dzugutov, Phys. Rev. B {\bf 65}, 104302 (2002).

\bibitem{mct} W. G\"otze and M. R. Mayr, Phys. Rev. E {\bf 61}, 587
(2000); T. Voigtmann, J. Non-Cryst. Solids {\bf 307-310}, 188 (2002).

\bibitem{ermZee} M. M\'ezard, G. Parisi and A. Zee, Nucl. Phys. B {\bf
559}, 689 (1999).

\bibitem{erm2} V. Mart\'\i{}n-Mayor, M. M\'ezard, G. Parisi and
P. Verrocchio, J. Chem. Phys. {\bf 114}, 8068 (2001).

\bibitem{erm2bis} T. S. Grigera, V. Mart\'\i{}n-Mayor, G. Parisi, and
P. Verrocchio, Phys. Rev. Lett. {\bf 87,} 085502 (2001).

\bibitem{metha} M. L. Mehta, {\em Random Matrices}, Academic Press,
London (1991).

\bibitem{erm} T. M. Wu and R. F. Loring, J. Chem. Phys. {\bf 97}, 8568
(1992); Y. Wan and R. Stratt, J. Chem. Phys. {\bf 100}, 5123 (1994);
A. Cavagna, I. Giardina and G. Parisi, Phys. Rev. Lett. {\bf 83,} 108
(1999).

\bibitem{bpvsheat} A. P. Sokolov, U. Buchenau, W. Steffen,
B. Frick,A. Wischnewski, Phys. Rev. B {\bf 52}, 9815 (1995);
N. J. Tao, G. Li, X. Chen, W. M. Du, H. Z. Cummins, Phys. Rev. A {\bf
44}, 6665 (1991); D. Engberg, A. Wischnewski, U. Buchenau,
L. B\"orjesson, A. J. Dianoux, A. P. Sokolov, L. M. Torell
Phys. Rev. B {\bf 59}, 4053 (1999).

\bibitem{bpsilica1} A. Wischnewski, U. Buchenau, A. J. Dianoux,
W. A. Kamitakahara, J. L. Zarestky, Phys. Rev. B {\bf 57}, 2663 (1998).

\bibitem{bpsilica2} S. Sugai and A. Onodera, Phys. Rev. Lett. {\bf 77}
4210 (1996); Y. Inamura, M. Arai, N. Kitamura, S. M. Bennington,
A. C. Hannon, Physica B {\bf 241-243}, 903 (1998); Y. Inamura,
M. Arai, O. Yamamuro, A. Inaba, N. Kitamura, T. Otomo, T. Matsuo,
S. M. Bennington, A. C. Hannon, Physica B {\bf 263-264}, 299 (1999);
P. Jund and R. Jullien, J. Chem. Phys. {\bf 113}, 2768 (2000).

\bibitem{soft-potential} V. G. Karpov et al., Sov. Phys. JETP {\bf
57,} 439 (1983); U. Buchenau, Yu. M. Galperin, and V. L. Gurevich,
Phys. Rev. B {\bf 46,} 2798 (1992).

\bibitem{ruocco00} G. Ruocco, F. Sette, R. Di Leonardo, G. Monaco,
M. Sampoli, T. Scopigno, and G. Viliani, Phys. Rev. Lett. {\bf 84},
5788 (2000).

\bibitem{horbach99} J. Horbach, W. Kob, and K. Binder,
J. Phys. Chem. B {\bf 103}, 4104 (1999).

\bibitem{pilla02} O. Pilla, S.Caponi, A.Fontana, M.Montagna, F.Rossi,
G.Viliani, L. Angelani, G. Ruocco, G. Monaco, F. Sette,
cond-mat/0209519.

\bibitem{iofferegel} A. F. Ioffe and A. R. Regel, Prog. Semicond. {\bf
4}, 237 (1960); S. N. Taraskin and S. R. Elliott, Phys. Rev. {\bf B
61}, 12017 (2000).

\bibitem{alexander89} S. Alexander, Phys. Rev. B {\bf 40,} 7953
(1989).

\bibitem{courtens96} see Foret et al. in \cite{debate}.

\bibitem{diffusoni} J. Fabian and P. B. Allen, Phys. Rev. Lett. {\bf
77}, 3839 (1996); J. L. Feldman, P. B. Allen, and S. R. Bickham,
Phys. Rev. B {\bf 59}, 3551 (1999); P. B. Allen, J. L. Feldman,
J. Fabian, F. Wooten, Phil. Mag. B, {\bf 82} (2002).

\bibitem{erm3} T. S. Grigera, V. Martin-Mayor, G. Parisi, and
P. Verrocchio, J. Phys.: Cond. Matt. {\bf 14}, 2167 (2002).

\bibitem{scopigno02} T. Scopigno, E. Pontecorvo, R. Di Leonardo,
M. Krisch, G. Monaco, G. Ruocco, B. Ruzicka, F. Sette,
cond-mat/0205615.

\bibitem{nature} T. S. Grigera, V. Mart\'\i{}n-Mayor, G. Parisi, and
P. Verrocchio, Nature {\bf 422}, 289 (2003).

\bibitem{saddles} L. Angelani, R. Di Leonardo, G. Ruocco, A. Scala,
and F. Sciortino, Phys. Rev. Lett. {\bf 85}, 5356 (2000); K. Broderix,
K. K. Bhattacharya, A. Cavagna, A. Zippelius, and I. Giardina,
Phys. Rev. Lett. {\bf 85}, 5360 (2000); T. S. Grigera, A. Cavagna,
I. Giardina, and G. Parisi, Phys. Rev. Lett. {\bf 88}, 055502 (2002);

\bibitem{mom} C. Benoit, E. Royer and G Poussigue, J. Phys.:
Condens. Matter {\bf 4}, 3125 (1992); P. Turchi, F. Ducastelle
and G. Treglia, J. Phys. C 15, 2891 (1982). 

\bibitem{INM} A. Rahman, M. Mandell, and J.P. McTague,
J. Chem. Phys. {\bf 64}, 1564 (1976); R. Cotterill, and U. Masden,
Phys. Rev. B {\bf 33}, 262 (1986); G. Seeley and T. Keyes,
J. Chem. Phys. {\bf 91}, 5581 (1989); B. Madan and T. Keyes,
J. Chem. Phys. {\bf 98}, 3342 (1992); T. Keyes, J. Chem. Phys. {\bf
101}, 5081 (1994); M. Cho, G.R. Fleming, S. Saito, I. Ohmine, and
R.M. Stratt, J. Chem. Phys. {\bf 100}, 6672 (1994); S. Bembenek and
B. Laird, Phys. Rev. Lett. {\bf 74}, 936 (1995); S. Bembenek and
B. Laird, J. Chem. Phys. {\bf 104}, 5199 (1996); T. Keyes,
J. Chem. Phys. A {\bf 101}, 2921 (1997).

\bibitem{workinprogress} Actually, a small number of terms in the
perturbative expansion are not obtained correctly in this way, and the
cactus equation is slightly modified~\cite{workinprogress2}.  However,
the results of sec.~III are unchanged.

\bibitem{workinprogress2} S. Ciliberti, T. S. Grigera,
V. Martin-Mayor, G. Parisi, P. Verrocchio, in preparation.

\bibitem{HYPERQUENCH} C. A. Angell, Y. Yue, L.-M. Wang,
J. R. D. Copley, S. Borick and S. Mossa, J.  Phys.: Condens. Matter
{\bf 15,} S1051 (2003).

\end{thebibliography}
\end{document}